\documentclass{IEEEoj}
\usepackage{cite}
\usepackage{amsmath,amssymb,amsfonts}
\usepackage{algorithm}
\usepackage{algpseudocode}
\usepackage{graphicx,color}
\usepackage{textcomp}
\usepackage{booktabs}
\usepackage{array}
\usepackage[hidelinks]{hyperref}
\usepackage{acronym} 
\def\BibTeX{{\rm B\kern-.05em{\sc i\kern-.025em b}\kern-.08em
    T\kern-.1667em\lower.7ex\hbox{E}\kern-.125emX}}
\AtBeginDocument{\definecolor{ojcolor}{cmyk}{0.93,0.59,0.15,0.02}}
\def\OJlogo{\vspace{-4pt}\hskip-4pt\includegraphics[height=18pt]{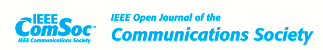}}

\acrodef{vlc}[VLC]{visible light communication}
\acrodef{led}[LED]{light emitting diode}
\acrodef{irs}[IRS]{intelligent reflecting surface}
\acrodef{nlos}[NLoS]{non-line-of-sight}
\acrodef{los}[LoS]{line-of-sight}
\acrodef{cir}[CIR]{channel impulse response}
\acrodef{cfr}[CFR]{channel frequency response}
\acrodef{snr}[SNR]{signal-to-noise ratio}
\acrodef{pls}[PLS]{physical layer security}
\acrodef{isi}[ISI]{inter-symbol interference}
\acrodef{dco-ofdm}[DCO-OFDM]{direct current optical orthogonal frequency division multiplexing}
\acrodef{mrc}[MRC]{maximal-ratio combining}
\acrodef{pd}[PD]{photodetector}
\acrodef{rl}[RL]{reinforcement learning}
\acrodef{mdp}[MDP]{Markov decision process}
\acrodef{ppo}[PPO]{proximal policy optimisation}
\acrodef{csi}[CSI]{channel state information}
\acrodef{rf}[RF]{radio frequency}
\acrodef{6g}[6G]{sixth generation}
\acrodef{fov}[FoV]{field of view}

\acrodef{lifi}[LiFi]{light fidelity}
\acrodef{iot}[IoT]{internet of things}
\acrodef{ofdm}[OFDM]{orthogonal frequency division multiplexing}
\acrodef{imdd}[IM/DD]{intensity modulation and direct detection}
\acrodef{noma}[NOMA]{non-orthogonal multiple access}
\acrodef{miso}[MISO]{multiple-input single-output}
\acrodef{an}[AN]{artificial noise}
\begin{document}
\receiveddate{XX Month, XXXX}
\reviseddate{XX Month, XXXX}
\accepteddate{XX Month, XXXX}
\publisheddate{XX Month, XXXX}
\currentdate{11 January, 2024}
\doiinfo{OJCOMS.2024.011100}
\title{ Enhancing PLS of Indoor IRS-VLC Systems for Colluding and Non-Colluding Eavesdroppers}

\author{Rashid Iqbal\IEEEauthorrefmark{1}\IEEEmembership{(Graduate Student Member, IEEE)}, 
Ahmed Zoha\IEEEauthorrefmark{1}\IEEEmembership{(Senior Member, IEEE)}, 
Salama Ikki\IEEEauthorrefmark{2}\IEEEmembership{(Senior Member, IEEE)}, 
Muhammad Ali Imran\IEEEauthorrefmark{1}\IEEEmembership{(Fellow, IEEE)}, 
and Hanaa Abumarshoud\IEEEauthorrefmark{1}\IEEEmembership{(Senior Member, IEEE)}}

\affil{James Watt School of Engineering, University of Glasgow, Glasgow G12 8QQ, U.K.}
\affil{Department of Electrical and Computer Engineering, Lakehead University, Thunder Bay, ON, Canada.}

\corresp{CORRESPONDING AUTHOR: Hanaa Abumarshoud (e-mail: Hanaa.Abumarshoud@glasgow.ac.uk).}

\begin{abstract}
Most intelligent reflecting surface (IRS)-aided indoor visible light communication (VLC) studies ignore the time delays introduced by reflected paths, even though these delays are inherent in practical wideband systems. In this work, we adopt a realistic assumption of IRS-induced time delay for physical layer security (PLS) enhancement. We consider an indoor VLC system where an IRS is used to shape the channel so that the reflected signals add constructively at the legitimate user and create intersymbol interference at eavesdroppers located inside the coverage area. The resulting secrecy capacity maximisation over the IRS element allocation is formulated as a complex combinatorial optimisation problem and is solved using deep reinforcement learning with proximal policy optimisation (PPO). The approach is evaluated for both colluding eavesdroppers, which combine their received signals, and non-colluding eavesdroppers, which act independently. Simulation results are presented for a range of indoor layouts and user placements, demonstrating significant secrecy capacity gains. In a worst-case scenario, where the eavesdroppers have stronger channels than the legitimate user, the proposed PPO-based IRS allocation improves secrecy capacity by 107\% and 235\% in the colluding and non-colluding cases, respectively, compared with allocating all IRS elements to the legitimate user. These results demonstrate that time-delay-based IRS control can provide a strong secrecy advantage in practical indoor VLC scenarios.
\end{abstract}


\maketitle

\section{INTRODUCTION}
\IEEEPARstart{V}{sible} light communication (VLC) has attracted significant interest as a key enabling technology for future \ac{6g} wireless networks. The rapidly growing demand for high speed wireless connectivity and the increasing congestion and licensing cost of the \ac{rf} spectrum motivate the search for alternative access technologies, and VLC addresses this demand by exploiting the large optical band between approximately 400~THz and 790~THz \cite{haas2016lifi}. In addition to this abundant bandwidth, VLC offers several attractive features, including licence free operation, high energy efficiency through \ac{led} luminaires, low electromagnetic interference, and the potential for seamless integration with existing solid state lighting infrastructure \cite{elgala2011indoor,karunatilaka2015ledvlcsurvey}. These properties make VLC particularly well suited to high density indoor environments such as offices, hospitals, and industrial facilities, where reliable, secure, and fast throughput connectivity is required.

Despite these advantages, security remains a critical concern for VLC systems. The optical signal is largely confined within the \ac{los} region and does not penetrate opaque walls, which provides a higher degree of spatial confinement compared with conventional \ac{rf} links. However, in typical indoor deployments, any user located within the illumination footprint of the \ac{led} luminaires can potentially intercept the transmitted signal, which exposes the system to unauthorised eavesdropping \cite{Arafa_2019,mostafa2015physicallayer}. Traditional security mechanisms that rely on upper layer cryptographic protocols are often not sufficient in this context, especially in dense networks with a large number of low power \ac{iot}, since key distribution, key renewal, and authentication procedures introduce non-negligible overhead and latency \cite{arfaoui2020vlc_pls_survey}. These limitations have motivated the use of \ac{pls} techniques, which exploit the randomness and spatial characteristics of the wireless channel to provide information theoretic security by ensuring that the achievable rate at the legitimate receiver, Bob, is strictly higher than that at the eavesdropper, Eve \cite{bloch2011wireless,sezgin2009secrecy}. The associated performance metric, known as the secrecy capacity, quantifies the maximum rate at which perfectly secure communication can be achieved.

Building upon the need for secure indoor communication, several works have investigated \ac{pls} techniques tailored to indoor VLC systems. In these studies, the inherent directionality of the optical beam and the geometric constraints of the room are exploited to enhance secrecy performance. Beamforming and null steering at the \ac{led}, often combined with artificial noise injection, have been proposed to improve the secrecy rate of multiple input single output wiretap channels in indoor scenarios \cite{mostafa2015physicallayer,pham2024AN_VLC}. In addition, optical elements such as lenses or multibeam structures have been employed to further confine the useful signal in space and to reduce information leakage towards potential eavesdroppers \cite{cirkinoğlu2021lens_pls,ding2022multibeam_pls}. Collectively, these approaches show that indoor VLC links can provide strong secrecy guarantees when the transmitter configuration and room geometry are carefully designed. However, the propagation environment is typically treated as fixed, and the opportunity to reconfigure the channel through additional programmable surfaces remains largely unexplored.

Recently, \ac{irs} technology has been introduced into VLC systems as a means to actively manipulate the indoor propagation environment and to overcome the limitations of fixed geometry. An \ac{irs} typically consists of a planar array of passive or low power reflecting elements whose optical reflection properties can be individually controlled, which allows the incident light to be redirected in a programmable manner \cite{aboagye2021irs_vlc,sun2022irs_resource}. In indoor VLC, such programmable surfaces can be deployed on walls or ceilings to create additional controllable paths between the transmitter and the users. From a \ac{pls} perspective, this provides a new degree of freedom, since appropriate configuration of the surface can concentrate the received power at the legitimate user while suppressing the signal at the eavesdropper location, which leads to a significant improvement in secrecy performance \cite{abumarshoud2023icc_pls,qian2024secureVLC_irs}.

However, most existing works on \ac{irs}-assisted VLC security rely on simplified channel models. A common assumption is that the composite channel formed by the \ac{los} link and the \ac{irs} reflected links is frequency flat, or that the temporal dispersion introduced by the reflected paths can be neglected \cite{aboagye2021irs_vlc,sun2022irs_resource,qian2024secureVLC_irs}. This assumption is reasonable for low rate narrowband modulation, yet it becomes inaccurate in practical high speed VLC systems that employ wideband schemes such as \ac{dco-ofdm}. In such systems, the path length differences between the direct \ac{los} component and the multiple \ac{nlos} components reflected by the \ac{irs} introduce non-negligible time delays, which give rise to frequency selective fading and \ac{isi}. These effects can degrade performance if they are not properly taken into account in the system and security analysis. A few recent channel modelling studies have started to characterise the impact of \ac{irs}-induced time delay on the frequency domain behaviour of VLC links \cite{chen2023freqdomainIRS,HAIDER2025102703}.

\subsection{Related Work}
\ac{pls} for indoor VLC was first studied mainly under the assumption that eavesdroppers act independently. Early works considered single cell scenarios in which one \ac{led} serves a single user in the presence of randomly located non-colluding eavesdroppers. In this setting, secrecy capacity and secrecy outage expressions were derived, and the impact of key parameters such as the transmit optical power, the receiver \ac{fov}, and the user locations was characterised \cite{pan2017secureVLC_random,cho2018secure_beamforming}. These results showed that indoor VLC can achieve strong secrecy performance thanks to the spatial confinement of the optical beam and the possibility of shaping the coverage region through power control and beamforming. Subsequent studies extended the analysis to more realistic LiFi networks with multiple access points and user mobility, and evaluated secrecy under imperfect \ac{csi} and device orientation, which provided a more accurate picture of practical deployments \cite{abumarshoud2021realistic_secrecy}. A variety of transmit techniques were proposed, including multiple input single output beamforming and artificial noise, power allocation strategies, and shaped optical beams, all aimed at improving the secrecy rate when eavesdroppers are non-colluding and experience independent channels \cite{arfaoui2020vlc_pls_survey,cho2018secure_beamforming}. 

The assumption that eavesdroppers do not cooperate may, however, be optimistic in dense indoor environments where multiple malicious users can share their observations. This has motivated a line of work on \ac{pls} with colluding eavesdroppers. In the context of VLC, \cite{cho2018colluding_vlc} analysed the secrecy performance of a downlink system with randomly located colluding eavesdroppers that combine their received signals, and derived closed form expressions for the secrecy outage probability using tools from stochastic geometry, while \cite{abumarshoud2021realistic_secrecy} considered a multi cell LiFi network and showed that colluding eavesdroppers can significantly reduce the achievable secrecy capacity, especially when realistic receiver orientation is taken into account. More broadly, the detrimental effect of collusion on secrecy capacity has been quantified in the wireless literature, where information theoretic analyses demonstrated that cooperation among eavesdroppers shrinks the security region and tightens secrecy constraints \cite{sezgin2009secrecy,bloch2011wireless}. Overall, these studies highlighted that indoor VLC is vulnerable to colluding eavesdroppers, and that system design based solely on non-colluding models may overestimate the available secrecy capacity.

\Ac{irs} have also been investigated in the context of \ac{pls}. By properly controlling the reflection properties of a large number of low cost elements, an \ac{irs} can strengthen the legitimate channel. Motivated by this concept, several works have introduced \ac{irs} concepts into indoor VLC. Programmable optical reflectors mounted on walls or ceilings have been used to create additional controllable paths between the transmitter and the receivers, and have been shown to improve coverage and enhance secrecy, for example in \ac{noma} based VLC systems and in downlink LiFi scenarios with non-colluding eavesdroppers \cite{abumarshoud2023icc_noma_irs_vlc,qian2024secureVLC_irs,abumarshoud2024irs_vlc_secure,HAIDER2025102703}. In most of these designs, the composite channel formed by the \ac{los} link and the \ac{irs} reflected links was modelled as frequency flat, and the propagation delay associated with the reflected paths was not explicitly taken into account. 

In practical wideband VLC systems that employ intensity modulation with direct detection, the different path lengths between the light source, the \ac{irs} elements, and the receivers introduce non-negligible time delays, which lead to frequency selective fading and \ac{isi}. Our recent work \cite{leveraging2024irs_delay_pls} was the first to exploit this \ac{irs} induced time delay as a degree of freedom for enhancing \ac{pls} in indoor VLC. In that work, a single eavesdropper was considered, and an optimisation problem was formulated to allocate the \ac{irs} elements in order to maximise the secrecy capacity by creating constructive \ac{isi} at the legitimate user and destructive \ac{isi} at the eavesdropper. The numerical results showed that a time delay aware use of the \ac{irs} can provide substantial secrecy gains compared to configurations that ignore the time delay. However, the analysis in \cite{leveraging2024irs_delay_pls} was limited to one eavesdropper, and did not address scenarios with multiple non-colluding and colluding eavesdroppers under a unified time delay \ac{irs} framework. To the best of our knowledge, a comparative secrecy capacity analysis for colluding and non-colluding eavesdroppers in time delay \ac{irs}-assisted VLC systems has not been reported so far.

\subsection{Motivation and Contributions}
In this paper, we use the \ac{irs} induced time delay to improve the \ac{pls} of an indoor VLC system. Rather than neglecting the time delay in VLC channels, we shape the channel impulse response so that the reflected signals add constructively at Bob and destructively at the eavesdroppers. By carefully choosing how the \ac{irs} elements are allocated, strong reflected paths align with the \ac{los} component at Bob, while becoming misaligned at the eavesdroppers, creating harmful \ac{isi} that degrades their reception. This effect is particularly pronounced when the eavesdroppers are located closer to the transmitter than Bob, making it a powerful tool in challenging geometric configurations. Furthermore, to strengthen the threat model, we consider a general scenario with \(K\) eavesdroppers that may act independently or collude by combining their received signals. However, the secrecy capacity maximisation problem that arises from this formulation involves discrete allocation of the \ac{irs} elements, rendering it non-convex and combinatorial in nature. To overcome this challenge, we employ a \ac{rl} method based on \ac{ppo}, which learns an effective allocation strategy for the \ac{irs} elements through direct interaction with the environment.
The main contributions of this paper are as follows:
\begin{itemize}
\item We formulate a secrecy capacity maximisation problem for an indoor VLC system assisted by an \ac{irs}, where the discrete allocation of the \ac{irs} elements shapes the channel frequency response to create constructive interference for Bob and destructive interference for both non-colluding and colluding eavesdroppers.
\item Moreover, we apply a \ac{ppo} based \ac{rl} algorithm with carefully designed state, action, and reward definitions, enabling the agent to learn an \ac{irs} allocation strategy that consistently improves long-term secrecy performance.
\item Finally, we present simulation results that illustrate the secrecy capacity behaviour of the proposed \ac{irs}-induced time-delay approach for varying numbers and locations of eavesdroppers, demonstrating that positive secrecy can be maintained even in worst-case geometric configurations where colluding eavesdroppers possess stronger channels than Bob.
\end{itemize}

\section{System Model}

\begin{figure}[t]
\centering
\centerline{\includegraphics[width=3.3in]{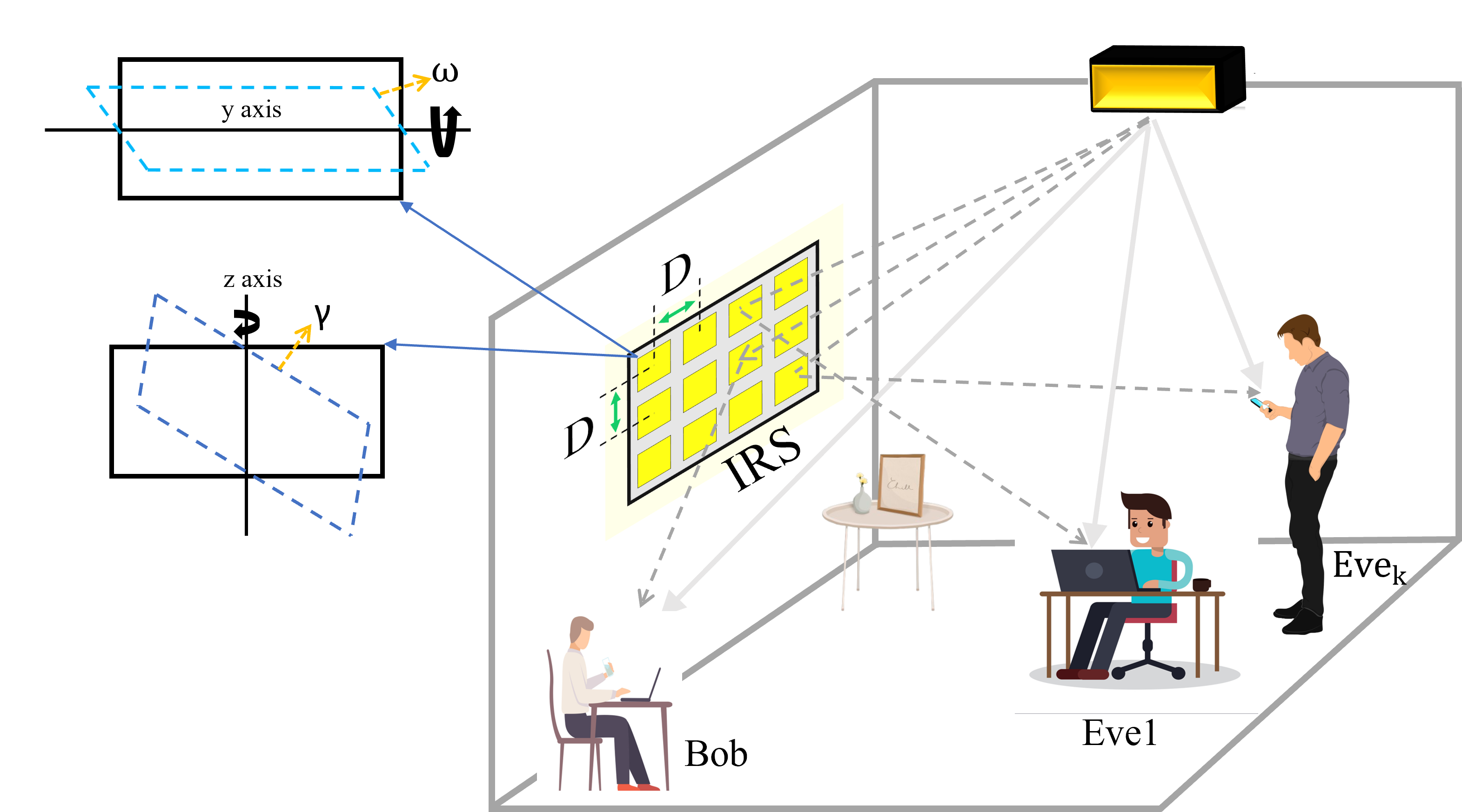}}
\caption{ System model of an indoor \ac{irs}-assisted VLC system with one legitimate user and $K$ eavesdropper.\color{black}}
\label{fig_a}
\end{figure}

As shown in Fig.~\ref{fig_a}, an indoor VLC scenario is considered. In this environment, a single \ac{led} transmitter, Alice, is mounted at the centre of the ceiling to provide simultaneous illumination and high-speed data transmission. The confidential message is intended exclusively for a single legitimate user, Bob; however, the widebeam nature of the \ac{led} light ensures that the signal covers a large footprint, allowing any receiver within this area to potentially intercept the transmission. Consequently, a security threat is modelled in the form of $K$ passive eavesdroppers, denoted by the set $\mathcal{E} = \{E_1,\dots,E_K\}$, who are arbitrarily located within the room. These eavesdroppers are assumed to remain physically silent to avoid detection by the transmitter.  Although these eavesdroppers do not reveal their presence through uplink transmissions, we assume that their \ac{csi} is perfectly known at the transmitter to investigate the fundamental secrecy limits, which follows the established analytical convention in \ac{pls} research to evaluate performance under a theoretical worst-case security benchmark. This assumption is supported by recent advancements in optical wireless channel estimation, such as the closed-form location and orientation estimation framework presented in \cite{Bozanis}, which demonstrates the practical feasibility of acquiring such detailed CSI in VLC systems.

To enhance the \ac{pls} of the legitimate link, an \ac{irs} is mounted on one of the walls. The \ac{irs} consists of \(N_{\text{irs}}\) passive reflecting elements with separation distance \(D\) between adjacent elements. The incident optical signal from the \ac{led} illuminates the \ac{irs}, and each element reflects part of this signal towards the interior of the room by adjusting its orientation. In particular, each element is oriented to steer its reflection toward a selected receiver direction, hence a single element is assumed to be allocated to one user at a time, in line with mirror array based optical IRS operation \cite{fd5}. In this manner, additional \ac{nlos} path is created between Alice and each receiver, with propagation distances that depend on the positions of the \ac{irs} elements and of the users. The different path lengths associated with these links give rise to different arrival times at the receivers and contribute to the overall time dispersion of the channel. In the remainder of this section, the end-to-end optical channel for each receiver is expressed as the superposition of the direct \ac{los} component and the \ac{irs}-assisted \ac{nlos} components, and the corresponding \ac{cir} and \ac{cfr} expressions used in the secrecy capacity analysis are derived.\color{black}

\setlength{\abovecaptionskip}{5pt}
\setlength{\abovecaptionskip}{22pt}
\subsection{LoS Channel}
The direct \ac{los} path in the VLC link is modelled using the modified Lambertian expression. For any user $k\in\{B,E_1,\dots,E_K\}$, the \ac{los} channel gain $g_{\scriptscriptstyle{k,\text{LED}}}^{\scriptscriptstyle \text{LOS}}$ is determined by the link geometry and the optical front end parameters, and is given by \color{black}
\begin{equation}
\begin{split}
g_{\scriptscriptstyle{k,\text{LED}}}^{\scriptscriptstyle \text{LOS}} =
\begin{cases}
\frac{(m+1)A}{2\pi d_{\scriptscriptstyle{k,\text{LED}}}^{2}}\,
\cos^{m}\!\bigl(\phi_{\scriptscriptstyle{k,\text{LED}}}\bigr)\,
\cos\!\bigl(\psi_{\scriptscriptstyle{k,\text{LED}}}\bigr)\,
\mathbb{G}_{\scriptscriptstyle f}\,\mathbb{G}_{\scriptscriptstyle c}\,,\\
\hspace{8em}\text{for } 0\le \psi_{\scriptscriptstyle{k,\text{LED}}}\le \Psi_{\scriptscriptstyle \text{FoV}}\,,\\
0\,,\hspace{10.5em}\text{otherwise}
\end{cases}
\end{split}
\label{eq1}
\end{equation}
where $A$ is the physical \ac{pd} area and $d_{\scriptscriptstyle{k,\text{LED}}}$ is the \ac{los} distance from Alice to user $k$. The model parameters include the \ac{led} irradiance angle $\phi_{\scriptscriptstyle{k,\text{LED}}}$, the receiver incidence angle $\psi_{\scriptscriptstyle{k,\text{LED}}}$, the optical filter gain $\mathbb{G}_{\scriptscriptstyle f}$, and the optical concentrator gain $\mathbb{G}_{\scriptscriptstyle c}$. The channel gain depends strongly on the geometry. The term $\cos^{m}(\phi)$ models the \ac{led} radiation pattern, while $\cos(\psi)$ accounts for the receiver orientation. The \ac{fov} constraint $\Psi_{\scriptscriptstyle \text{FoV}}$ imposes a hard cut off, since any ray arriving at an angle greater than $\Psi_{\scriptscriptstyle \text{FoV}}$ is rejected and leads to zero gain. The concentrator gain $\mathbb{G}_{\scriptscriptstyle c}$ is a function of the lens refractive index $\xi$, and $\Psi_{\scriptscriptstyle \text{FoV}}$ is expressed as \color{black}
\begin{equation}
\mathbb{G}_{c}=
\begin{cases}
\frac{\xi^{2}}{\sin^{2}\!\bigl(\Psi_{\scriptscriptstyle \text{FoV}}\bigr)}\,, & 0\le \psi_{\scriptscriptstyle{k,\text{LED}}}\le \Psi_{\scriptscriptstyle \text{FoV}}\,,\\
0\,, & \text{otherwise},
\end{cases}
\end{equation}
and the Lambertian order $m$ in \eqref{eq1} quantifies the directivity of the \ac{led} beam and is derived from the semi angle at half power $\Phi_{\scriptscriptstyle 1/2}$ as
\begin{equation}
m = -\frac{1}{\log_{2}\!\bigl(\cos(\Phi_{\scriptscriptstyle 1/2})\bigr)}.
\end{equation}
\color{black}
\subsection{NLoS Channel}
The \ac{nlos} channel is modelled by considering an \ac{irs} composed of $N_{\text{irs}}$ passive reflecting elements mounted on a wall. Each element exhibits reflectivity $\rho$ and acts as a specular reflector for the incident light from the \ac{led}. The specular reflection gain from the \ac{led} to element $n$ and then to user $k$, denoted by $g_{\scriptscriptstyle{k,n,\text{LED}}}^{\scriptscriptstyle \text{NLOS}}$, is written as
\begin{equation}
\begin{split}
g_{\scriptscriptstyle{k,n,\text{LED}}}^{\scriptscriptstyle \text{NLOS}} =
\begin{cases}
\frac{\rho\,(m+1)\,A}{2\pi\,(d_{\scriptscriptstyle{n,\text{LED}}}+d_{\scriptscriptstyle{k,n}})^{2}}\,
\cos^{m}\!\bigl(\phi_{\scriptscriptstyle{n,\text{LED}}}\bigr)\,
\cos\!\bigl(\psi_{\scriptscriptstyle{k,n}}\bigr)\,
\mathbb{G}_{\scriptscriptstyle f}\,\mathbb{G}_{\scriptscriptstyle c}\,,\\
\hspace{8em}\text{for } 0\le \psi_{\scriptscriptstyle{k,n}}\le \Psi_{\scriptscriptstyle \text{FoV}}\,,\\
0\,,\hspace{10.5em}\text{otherwise},
\end{cases}
\end{split}
\end{equation}
where $d_{\scriptscriptstyle{n,\text{LED}}}$ and $d_{\scriptscriptstyle{k,n}}$ denote the Euclidean distances from the \ac{led} to element $n$ and from element $n$ to user $k$, respectively. The angles $\phi_{\scriptscriptstyle{n,\text{LED}}}$ and $\psi_{\scriptscriptstyle{k,n}}$ represent the irradiance and incidence angles associated with the path via element $n$, and the gain is limited by the receiver \ac{fov}, $\Psi_{\scriptscriptstyle \text{FoV}}$, through the same filter and concentrator terms used in the \ac{los} case.

This work considers realistic assumptions and accounts for the time delay in light propagation for both \ac{los} and \ac{nlos} links. For user $k$, the propagation delay of the direct \ac{los} path is $\tau_{\scriptscriptstyle{k,\text{LED}}} = d_{\scriptscriptstyle{k,\text{LED}}}/c$, while the delay associated with the path via element $n$ is $\tau_{\scriptscriptstyle{k,n,\text{LED}}} = (d_{\scriptscriptstyle n,\text{LED}} + d_{\scriptscriptstyle k,n})/c$, where $c$ is the speed of light. To represent the overall channel observed by user $k$, the \ac{los} and \ac{nlos} contributions are combined into the \ac{cir} $q_k(t)$ as \cite{chen2023freqdomainIRS}
\begin{equation}
q_k(t)=g_{\scriptscriptstyle{k,\text{LED}}}^{\scriptscriptstyle \text{LOS}} \delta\!\bigl(t-\tau_{\scriptscriptstyle{k,\text{LED}}}\bigr)
+\sum_{n=1}^{N_{\scriptscriptstyle{\text{irs}}}} g_{\scriptscriptstyle{k,n,\text{LED}}}^{\scriptscriptstyle \text{NLOS}} \delta\!\bigl(t-\tau_{\scriptscriptstyle{k,n,\text{LED}}}\bigr),
\label{equation:cfr}
\end{equation}
where $\delta(\cdot)$ denotes the unit impulse function.

To examine how the \ac{irs} induced time delay affects the channel across frequency, the corresponding \ac{cfr} $Q_k(f)$ is used. It is obtained as the Fourier transform of the \ac{cir} in \eqref{equation:cfr} and can be written as
\begin{equation}
Q_k(f) = g_{\scriptscriptstyle{k,\text{LED}}}^{\scriptscriptstyle \text{LOS}}e^{-j2\pi f\tau_{\scriptscriptstyle{k,\text{LED}}}}
+ \sum_{n=1}^{N_{\scriptscriptstyle \text{irs}}} g_{\scriptscriptstyle{k,n,\text{LED}}}^{\scriptscriptstyle \text{NLOS}}e^{-j2\pi f\tau_{\scriptscriptstyle{k,n,\text{LED}}}}.
\label{equa:5}
\end{equation}
The \ac{snr} at frequency $f$ for user $k$ compares the power of the modulated signal to the noise power at the \ac{pd} and is written as \cite{chen2023freqdomainIRS}
\begin{equation}
\gamma_k(f) = \frac{2T_{\scriptscriptstyle s}E(f)P_{\scriptscriptstyle{\text{opt}}}^{2}\, |Q_k(f)|^{2} R_{\scriptscriptstyle{\text{pd}}}^{2}}{\Gamma \mathsf{k}^{2} \mathcal{N}},
\label{eq7}
\end{equation}
where $T_{\scriptscriptstyle s}$ is the symbol period, $E(f)$ is the transmit power distribution, $P_{\scriptscriptstyle{\text{opt}}}$ is the optical power transmitted by the \ac{led}, $R_{\scriptscriptstyle{\text{pd}}}$ is the \ac{pd} responsivity and $\mathcal{N}$ is the noise power spectral density at the \ac{pd}. The parameter $\Gamma$ is the gap factor and accounts for implementation losses due to modulation and coding, while the modulation scaling factor $\mathsf{k}$ is used to keep the signal amplitude within the operational range of the \ac{led}.

It follows from \eqref{equa:5} that the \ac{snr} depends on the channel power gain $|Q_k(f)|^{2}$. By writing $|Q_k(f)|^{2} = Q_k(f)Q_k^{*}(f)$ and grouping terms, the magnitude squared of the \ac{cfr} can be expressed as
\begin{equation}
\begin{small}
\begin{aligned}
|Q_k(f)|^{2}
&= \bigl(g^{\scriptscriptstyle \text{LOS}}_{\scriptscriptstyle k,\text{LED}}\bigr)^{2}
+ 2\,g^{\scriptscriptstyle \text{LOS}}_{\scriptscriptstyle k,\text{LED}}
\sum_{i=1}^{N_{\scriptscriptstyle \text{irs}}}
g^{\scriptscriptstyle \text{NLOS}}_{\scriptscriptstyle k,i,\text{LED}}
\cos\!\bigl(2\pi f\,\Delta\tau_{i,k}\bigr)+ \\&\sum_{i=1}^{N_{\scriptscriptstyle \text{irs}}}
\bigl(g^{\scriptscriptstyle \text{NLOS}}_{\scriptscriptstyle k,i,\text{LED}}\bigr)^{2}
+ \sum_{i=1}^{N_{\scriptscriptstyle \text{irs}}}\sum_{u<i}
2\,g^{\scriptscriptstyle \text{NLOS}}_{\scriptscriptstyle k,i,\text{LED}}
g^{\scriptscriptstyle \text{NLOS}}_{\scriptscriptstyle k,u,\text{LED}}
\cos\!\bigl(2\pi f\,\Delta\tau_{u,i,k}\bigr),
\label{fre_k}
\end{aligned}
\end{small}
\end{equation}
where $\Delta\tau_{i,k}=\tau_{\scriptscriptstyle k,i,\text{LED}}-\tau_{\scriptscriptstyle k,\text{LED}}$ and $\Delta\tau_{u,i,k}=\tau_{\scriptscriptstyle k,u,\text{LED}}-\tau_{\scriptscriptstyle k,i,\text{LED}}$. The first and third terms in \eqref{fre_k} are independent of frequency and represent the individual \ac{los} and \ac{nlos} power contributions, while the two cosine sums describe how the relative delays lead to constructive and destructive combining between the \ac{los} path and each \ac{nlos} path and among the \ac{nlos} paths themselves.

\section{Secrecy Performance Analysis}

In this section, the secrecy performance of the considered \ac{irs}-assisted indoor VLC system is analysed. Two eavesdropping scenarios are considered, namely non-colluding eavesdroppers and colluding eavesdroppers that combine their received signals. Due to the different path lengths of the \ac{los} and \ac{nlos} links, the \ac{irs} induced time delays lead to different channel frequency responses for Bob and for the eavesdroppers. These differences are reflected in the achievable rates of the users and, consequently, in the secrecy capacity.

The starting point of the analysis is the achievable rate of a generic user $k$ in the \ac{dco-ofdm} link. Under the information theoretic model for a frequency selective channel, the rate $R_k$ is obtained by integrating over frequency \cite{chen2023freqdomainIRS}
\begin{equation}
\begin{aligned}
R_{\scriptscriptstyle k} &= \int_0^{\frac{1}{2 T_{\scriptscriptstyle s}}} \log _2\left(1+\gamma_k(f)\right) df.
\end{aligned}
\label{rate}
\end{equation}
 In the following two subsections, the rate is used to define the secrecy capacity for non-colluding and colluding eavesdroppers.

\subsection{Secrecy Capacity for Non-Colluding Eavesdroppers}

In this subsection, the secrecy capacity in the presence of non-colluding eavesdroppers is characterised. The $K$ eavesdroppers are assumed to act independently and do not share their received signals. In this case, the secrecy performance is limited by the eavesdropper with the largest achievable rate, and the secrecy capacity is written as
\begin{equation}
C_s^{\text{non-coll}} = \max\left(0, R_B - \max_{j \in \{1..K\}} \{ R_{E,j} \} \right),
\label{eq:cs_non_coll}
\end{equation}
where $R_B$ denotes the achievable rate at Bob and $R_{E,j}$ denotes the achievable rate at the $j$th eavesdropper.

The exact evaluation of \eqref{eq:cs_non_coll} requires the rate expression in \eqref{rate} to be computed for Bob and for all $K$ eavesdroppers. Since this expression involves the frequency selective \ac{snr} in \eqref{eq7} together with the channel power gain in \eqref{fre_k}, a closed-form solution is not available. To obtain a tractable expression and to highlight the role of the delays, a closed-form approximation for the rate $R_k$ is derived. 

For notational convenience, the \ac{los} gain is denoted by $G_{\text{los},k} \triangleq g_{\scriptscriptstyle{k,\text{LED}}}^{\scriptscriptstyle \text{LOS}}$, the $i$th \ac{nlos} gain by $G_{i,k} \triangleq g_{\scriptscriptstyle{k,i,\text{LED}}}^{\scriptscriptstyle \text{NLOS}}$, the \ac{los} delay by $\tau_{\text{los},k} \triangleq \tau_{\scriptscriptstyle{k,\text{LED}}}$, and the $i$th \ac{nlos} delay by $\tau_{i,k} \triangleq \tau_{\scriptscriptstyle{k,i,\text{LED}}}$. The frequency independent \ac{snr} prefix is written as $\Lambda = \frac{2T_{\scriptscriptstyle s}P_{\scriptscriptstyle \text{opt}}^{2}R_{\scriptscriptstyle \text{pd}}^{2}}{\Gamma\,\mathsf{k}^{2}\,\mathcal{N}}$, where a flat power distribution $E(f)=1$ is assumed. By substituting \eqref{fre_k} into \eqref{eq7} and then into \eqref{rate}, the rate integral for user $k$ can be written as
\begin{equation}
\begin{aligned}
R_k &= \int_{0}^{\frac{1}{2T_{\scriptscriptstyle s}}} \log_2 \Bigg( 1 + \Lambda \Bigg( G_{\text{los},k}^2 + \sum_{i=1}^{N_{\text{irs}}} G_{i,k}^2 \\
&\quad + 2 G_{\text{los},k} \sum_{i=1}^{N_{\text{irs}}} G_{i,k} \cos\bigl(2\pi f(\tau_{i,k} - \tau_{\text{los},k})\bigr) \\
&\quad + \sum_{i=1}^{N_{\text{irs}}} \sum_{u < i} 2 G_{i,k} G_{u,k} \cos\bigl(2\pi f(\tau_{u,k} - \tau_{i,k})\bigr) \Bigg) \Bigg) df.
\end{aligned}
\label{eq:Rk_expanded}
\end{equation}
To simplify this expression, we separate the frequency-independent term from the frequency–dependent term. The frequency–independent part is collected in
\begin{equation}
D_k = 1 + \Lambda \left(G_{\text{los},k}^2 + \sum_{i=1}^{N_{\text{irs}}} G_{i,k}^2 \right),
\label{eq:Dk_def}
\end{equation}
and the frequency–dependent part is written as
\begin{align}
N_k(f) = &\, 2G_{\text{los},k} \sum_{i=1}^{N_{\text{irs}}} G_{i,k} \cos\bigl(2\pi f (\tau_{i,k} - \tau_{\text{los},k})\bigr) \notag \\
& + \sum_{i=1}^{N_{\text{irs}}} \sum_{u < i} 2G_{i,k} G_{u,k} \cos\bigl(2\pi f (\tau_{u,k} - \tau_{i,k})\bigr).
\label{eq:Nk_def}
\end{align}
Using the identity $\log_2(1+x+y) = \log_2(1+x) + \log_2\!\bigl(1 + \frac{y}{1+x}\bigr)$, the rate in \eqref{eq:Rk_expanded} can be expressed as
\begin{equation}
R_k = \underbrace{\frac{1}{2T_{\scriptscriptstyle s}} \log_2 \left( D_k \right)}_{R_{k,1}}
+ \underbrace{\int_{0}^{\frac{1}{2T_{\scriptscriptstyle s}}} \log_2 \left( 1 + \frac{\Lambda N_k(f)}{D_k} \right) df}_{R_{k,2}},
\label{eq:rate_split_final}
\end{equation}
where $R_{k,1}$ represents the contribution of the frequency independent power and $R_{k,2}$ represents the additional contribution due to frequency selectivity.

The integral term $R_{k,2}$ involves the logarithm of a sum of cosine functions and is not available in closed form. To obtain an analytical approximation, the following steps are used. First, the cosine terms in $N_k(f)$ are expressed in terms of the delay differences and a balanced approximation for $\cos(x)$ based on $\frac{\sin(x)}{x}$ is employed, inspired by the inequality in \cite{Abramowitz1964}, which establishes that $\cos(x) \leq \frac{\sin(x)}{x} \leq 1$ for $0 \leq x \leq \pi$. Second, the main contribution of the integral is assumed to come from the dominant terms over the bandwidth $[0,f_{\text{max}}]$ with $f_{\text{max}} = \frac{1}{2T_s}$. This procedure leads to the following approximation for the total rate $R_k$
 \begin{align}
&R_k^{\text{approx}} \approx \frac{1}{2T_{\scriptscriptstyle s}} \log_2 \left(1 + \Lambda \left(G_{\text{los},k}^2 + \sum_{i=1}^{N_{\text{irs}}} G_{i,k}^2 \right) \right) \notag \\
&+ \frac{\Lambda}{D_k \ln(2)} \left( \frac{G_{\text{los},k}}{T_{\scriptscriptstyle s}} \sum_{i=1}^{N_{\text{irs}}} G_{i,k} \left(1 - \frac{\sin\left(\frac{\pi (\tau_{i,k} - \tau_{\text{los},k})}{T_{\scriptscriptstyle s}}\right)}{\frac{\pi (\tau_{i,k} - \tau_{\text{los},k})}{T_{\scriptscriptstyle s}}} \right) \right. \notag \\
&\left. + \frac{1}{T_{\scriptscriptstyle s}} \sum_{i=1}^{N_{\text{irs}}} \sum_{u < i} G_{i,k} G_{u,k} \left(1 - \frac{\sin\left(\frac{\pi (\tau_{u,k} - \tau_{i,k})}{T_{\scriptscriptstyle s}}\right)}{\frac{\pi (\tau_{u,k} - \tau_{i,k})}{T_{\scriptscriptstyle s}}} \right) \right).
\label{equa:approx_k}
\end{align}   
An approximated secrecy capacity for the non-colluding scenario is then obtained by applying \eqref{equa:approx_k} to Bob and to each of the $K$ eavesdroppers
\begin{equation}
C_s^{\text{non-coll}} \approx \max\left(0, R_B^{\text{approx}} - \max_{j \in \{1..K\}} \{ R_{E,j}^{\text{approx}} \} \right),
\label{eq:final_cs_non_coll}
\end{equation}
where $R_B^{\text{approx}}$ is obtained by setting $k=B$ in \eqref{equa:approx_k} and $R_{E,j}^{\text{approx}}$ is obtained by setting $k=E_j$.

\subsection{Secrecy Capacity for Colluding Eavesdroppers}

In this subsection, we consider the case where the $K$ eavesdroppers collude. The eavesdroppers are assumed to share their received signals and cooperate, so that their joint observation is more informative than any individual signal. This represents a stronger and more realistic threat model for dense indoor environments, and it is therefore important to characterise the secrecy capacity under this condition. The secrecy capacity is defined as the difference between Bob's achievable rate $R_B$ and the achievable rate of the eavesdroppers, which in this case is the rate of the colluding group, $R_E^{\text{coll}}$
\begin{equation}
C_s^{\text{coll}} = \max\left(0, R_B - R_E^{\text{coll}}\right).
\label{eq:cs_def}
\end{equation}

Bob's rate $R_B$ is obtained directly from the exact integral in \eqref{rate} by using his frequency dependent \ac{snr} $\gamma_B(f)$
\begin{equation}
R_B = \int_{0}^{\frac{1}{2T_{\scriptscriptstyle s}}} \log_2\left(1 + \gamma_B(f)\right) \, df.
\label{eq:rate_bob_new}
\end{equation}
For the colluding eavesdroppers, cooperation is modelled by \ac{mrc}, which provides an upper bound on their capability since it corresponds to coherent combining of their signals. With \ac{mrc}, the effective \ac{snr} of the group, $\gamma_E^{\text{coll}}(f)$, is the sum of the individual \acp{snr}
\begin{equation}
\gamma_E^{\text{coll}}(f) = \sum_{j=1}^{K} \gamma_{E_j}(f) = \sum_{j=1}^{K} \left( \frac{2T_{\scriptscriptstyle s}E(f)P_{\scriptscriptstyle \text{opt}}^{2}\,|Q_{E_j}(f)|^{2}R_{\scriptscriptstyle \text{pd}}^{2}}{\Gamma\,\mathsf{k}^{2}\,\mathcal{N}} \right).
\end{equation}
Assuming a constant $E(f)=1$ and using the \ac{snr} prefix $\Lambda$, this expression can be written as
\begin{equation}
\gamma_E^{\text{coll}}(f) = \Lambda \sum_{j=1}^{K} |Q_{E_j}(f)|^{2}.
\label{eq:snr_eve_coll_new}
\end{equation}
The achievable rate for the colluding eavesdroppers, $R_E^{\text{coll}}$, is then given by

 \begin{equation}
 \begin{small}
R_E^{\text{coll}} = \int_{0}^{\frac{1}{2T_{\scriptscriptstyle s}}} \log_2\left(1 + \Lambda \sum_{j=1}^{K} |Q_{E_j}(f)|^{2}\right) \, df.
\label{eq:rate_eve_coll_new}
\end{small}
\end{equation}   
Substituting \eqref{eq:rate_bob_new} and \eqref{eq:rate_eve_coll_new} into \eqref{eq:cs_def} and combining the two integrals yields
\begin{equation}
\begin{small}
C_s^{\text{coll}} = \left[ \int_{0}^{\frac{1}{2T_{\scriptscriptstyle s}}} \left( \log_2\left(1 + \gamma_B(f)\right) - \log_2\left(1 + \gamma_E^{\text{coll}}(f)\right) \right) df \right]^{+}.
\end{small}
\end{equation}    
Using the identity $\log(a) - \log(b) = \log(a/b)$, this expression can be rewritten as
\begin{equation}
C_s^{\text{coll}} = \left[ \int_{0}^{\frac{1}{2T_{\scriptscriptstyle s}}} \log_2\left( \frac{1 + \Lambda \, |Q_B(f)|^{2}}{1 + \Lambda \sum_{j=1}^{K} |Q_{E_j}(f)|^{2}} \right) df \right]^{+},
\label{eq:final_cs_colluding}
\end{equation}
where $[x]^{+} = \max(0, x)$. This integral form shows that a positive secrecy capacity is achieved when the frequency dependent \ac{snr} of Bob remains sufficiently larger than the combined \ac{snr} of the colluding eavesdroppers over the signal bandwidth, which in this work is controlled by the time delay configuration of the \ac{irs}.

\section{Optimisation and Learning Framework for IRS-Aided Secrecy Enhancement}

In this section, we present the optimisation framework used to improve the secrecy performance of the IRS-assisted system. We first describe how the secrecy capacity maximisation problem is formed by using a discrete model for the IRS element allocation. This problem becomes very large and difficult to solve by normal optimisation methods. For this reason, we then introduce a reinforcement-learning solution that can find a good IRS allocation without checking all possible combinations.

\subsection{Secrecy Capacity Maximisation Problem}
The channel model developed in the previous section is now used to formulate an optimisation problem, where the \ac{irs} elements are exploited to enhance the secrecy performance. As specified in the system model, the $N_{\text{irs}}$ elements are passive reflectors mounted on one wall, so their physical positions fix the possible path lengths and, hence, the associated delays. What can be configured is the allocation of each element towards a specific user, and each allocation pattern leads to a different set of effective delays at the receivers. To translate this physical capability into a tractable optimisation variable, a discrete allocation model is adopted in which each element $n$ is configured to serve exactly one user $k \in \{B, E_1, \dots, E_K\}$.

In this discrete allocation model, each \ac{irs} element is associated with exactly one user at a time. When an element is assigned to user $k$, its delay is adjusted so that its reflected component aligns constructively with the \ac{los} path at the receiver of user $k$. Because the users occupy different positions in the room, the propagation distances of the reflected paths are different for each user. As a result, an \ac{irs} allocation  that produces constructive combining for user $k$ does not in general produce constructive combining for users $j \neq k$, and tends instead to create non-constructive combining and additional \ac{isi} at their receivers. In this way, the discrete allocation captures the trade off between concentrating signal energy at Bob and dispersing it at the eavesdroppers.

The \ac{irs} configuration is represented by the discrete allocation vector
\begin{equation}
    \mathbf{a} = [a_1, \dots, a_{N_{\text{irs}}}],
\end{equation}
where $a_n \in \{B, E_1, \dots, E_K\}$ denotes the user served by the $n$th element. The vector $\mathbf{a}$ therefore represents a complete partition of the \ac{irs} elements among all users. Under this model, the \ac{cfr} for user $k$ becomes an explicit function of the allocation vector $\mathbf{a}$. It is assumed that only the elements allocated to user $k$ contribute coherently to its \ac{nlos} component, so that $Q_k(f)$ in \eqref{equa:5} is rewritten as
\begin{equation}
Q_k(f, \mathbf{a}) = g_{\scriptscriptstyle{k,\text{LED}}}^{\scriptscriptstyle \text{LOS}}e^{-j2\pi f\tau_{\scriptscriptstyle{k,\text{LED}}}}
+ \sum_{n \in \mathcal{N}_k(\mathbf{a})} g_{\scriptscriptstyle{k,n,\text{LED}}}^{\scriptscriptstyle \text{NLOS}}
e^{-j2\pi f\tau_{\scriptscriptstyle{k,n,\text{LED}}}},
\label{eq:cfr_with_a}
\end{equation}
where $\mathcal{N}_k(\mathbf{a}) = \{n \mid a_n = k\}$ denotes the set of element indices allocated to user $k$.

The objective of this work is to find the optimal IRS allocation vector $\mathbf{a}^*$ that maximises the colluding secrecy capacity. Mathematically, the optimisation problem $\mathcal{P}_1$ is formulated as follows:
\begin{small}
\begin{subequations}
\label{eq:opt_problem_block}
\begin{align}
\mathcal{P}_1: \max_{\mathbf{a}\in \mathcal{A}} \quad & C_s^{\text{coll}}(\mathbf{a}) \label{eq:obj_func} \\
\text{s.t.} \quad & a_n \in \{B, E_1, \ldots, E_K\}, \forall n \in \{1, \ldots, N_{\text{ins}}\}, \label{eq:c1} \\
& \sum_{k \in \{B, E_1, \ldots, E_K\}} \sum_{n=1}^{N_{\text{ins}}} \mathbb{I}(a_n = k) = N_{\text{ins}}, \label{eq:c2}
\end{align}
\end{subequations}
\end{small}
\\
where the objective function $C_s^{\text{coll}}(\mathbf{a})$ represents the secrecy capacity for the colluding scenario, derived by substituting the \acp{cfr} into the rate equation:
\begin{equation}
C_s^{\text{coll}}(\mathbf{a}) = \left[ \int_{0}^{\frac{1}{2T_{\scriptscriptstyle s}}} \log_2\left( \frac{1 + \Lambda \, |Q_B(f, \mathbf{a})|^{2}}{1 + \Lambda \sum_{j=1}^{K} |Q_{E_j}(f, \mathbf{a})|^{2}} \right) df \right]^{+}.
\label{eq:final_cs_colluding_a}
\end{equation}
In this formulation, $\mathcal{A}$ denotes the set of all admissible allocation vectors. Constraint \eqref{eq:c1} defines the discrete search space, enforcing that each IRS element $n$ must be allocated to exactly one user from the set $\{B, E_1, \dots, E_K\}$. Constraint \eqref{eq:c2}, utilising the indicator function $\mathbb{I}(\cdot)$, explicitly guarantees that the summation of allocated elements across all $K+1$ users equals the total number of IRS elements, $N_{\text{irs}}$, thereby ensuring a complete partition of the IRS array.

The problem $\mathcal{P}_1$ is a high dimensional, non-convex and combinatorial optimisation problem. The objective function in \eqref{eq:final_cs_colluding_a} is a non-linear and non-differentiable function of the discrete vector $\mathbf{a}$. Moreover, evaluating $C_s^{\text{coll}}(\mathbf{a})$ for a given allocation requires computing $K+1$ \acp{cfr} and performing the integral in \eqref{eq:final_cs_colluding_a}. The size of the search space $\mathcal{A}$ grows as $(K+1)^{N_{\text{irs}}}$, which makes exhaustive search infeasible for practical \ac{irs} sizes, for example $N_{\text{irs}}=100$. These complexity considerations motivate the use of a more scalable optimisation approach based on \ac{rl} in the next subsection.

\subsection{Proposed PPO Based Solution}

To solve the complex combinatorial optimisation problem $\mathcal{P}_1$, we reformulate it as a sequential decision making process and adopt a solution based on \ac{rl}. Specifically, the \ac{irs} allocation task is modelled as a \ac{mdp} and the \ac{ppo} algorithm is employed. \Ac{ppo} is a state-of-the-art policy gradient \ac{rl} method that is known for its data efficiency, ease of implementation, and robust stability in large discrete action spaces. These properties make it a suitable candidate for obtaining a high quality solution to $\mathcal{P}_1$ in a computationally feasible manner.

The components of the \ac{mdp} are defined as follows.
\begin{itemize}
    \item \textbf{Agent.} The \ac{rl} agent is the \ac{irs} controller, which must learn an allocation policy for the reflecting elements.
    
    \item \textbf{State $s_t$.} The process is modelled as a sequential allocation over $N_{\text{irs}}$ decision steps. At decision step $t \in \{1, \dots, N_{\text{irs}}\}$, the agent allocates the $t$th element. The state $s_t$ contains the allocation history up to that point and is written as $s_t = (a_1, a_2, \dots, a_{t-1})$. The initial state is $s_1 = \emptyset$. This representation allows the policy to capture dependencies between element allocations, where the decision for element $t$ can depend on earlier decisions.
    
    \item \textbf{Action $a_t$.} At each decision step $t$, the agent selects an action $a_t$ from the discrete action set $\mathcal{A}_t = \{B, E_1, \dots, E_K\}$. This action assigns the $t$th \ac{irs} element to one of the users and therefore determines to which user that element contributes coherent \ac{nlos} power.
    
    \item \textbf{Policy $\pi_\theta(a_t|s_t)$.} The agent's policy is represented by a deep neural network with parameters $\theta$. The policy network takes the state vector $s_t$ of dimension $N_{\text{irs}} + 9$ as input. This input consists of the allocation history combined with user coordinates. The network processes this input through a hidden layer with $d_h = 256$ neurons and ReLU activation. The output layer uses softmax activation to produce a probability distribution over $K+1$ possible actions. The action $a_t$ is sampled from this distribution. Training aims to adjust the parameters $\theta$ so that the resulting allocation sequences yield higher secrecy capacity.
    
    \item \textbf{Reward $R$.} The task is episodic. One episode consists of $N_{\text{irs}}$ decisions, which produce a complete allocation vector $\mathbf{a} = (a_1, \dots, a_{N_{\text{irs}}})$. After the last element has been allocated, the environment returns a terminal reward
    \begin{equation}
        R = C_s^{\text{coll}}(\mathbf{a}),
    \end{equation}
    where $C_s^{\text{coll}}(\mathbf{a})$ is given by \eqref{eq:final_cs_colluding_a}. All intermediate rewards $r_t$ for $t < N_{\text{irs}}$ are set to zero. This sparse reward structure means that the agent must learn the long term effect of its allocation sequence and directly optimises the secrecy capacity. The goal is to learn a policy $\pi_\theta$ that maximises the expected terminal reward $\mathbb{E}_{\mathbf{a} \sim \pi_\theta}[R]$, which corresponds to solving $\mathcal{P}_1$.
\end{itemize}

A separate value network with identical architecture is used to estimate the state value $V_\phi(s_t)$ for advantage computation. This critic network also has input dimension $N_{\text{irs}} + 9$ and a hidden layer with 256 neurons and ReLU activation. The output layer consists of a single linear neuron that produces the value estimate. The \ac{ppo} algorithm is used to optimise the policy $\pi_\theta$ by iteratively collecting experience and updating the network parameters. To improve training stability and efficiency, we collect a batch of multiple episodes before performing policy updates. Specifically, experience is accumulated over $N_{\text{batch}} = 64$ complete episodes before any network update is performed. Each trajectory in the batch is of the form
\[
\tau = (s_1, a_1, \dots, s_{N_{\text{irs}}}, a_{N_{\text{irs}}}, R),
\]
and is stored in a replay buffer $\mathcal{D}$. The policy is then updated by maximising the clipped surrogate objective
\begin{equation}
L^{\text{CLIP}}(\theta) = \hat{\mathbb{E}}_t \left[ \min\left( r_t(\theta) \hat{A}_t,\,
\text{clip}\bigl(r_t(\theta), 1-\epsilon, 1+\epsilon\bigr) \hat{A}_t \right) \right],
\end{equation}
where $r_t(\theta) = \frac{\pi_\theta(a_t|s_t)}{\pi_{\theta_{\text{old}}}(a_t|s_t)}$ is the probability ratio and $\hat{A}_t$ is an estimate of the advantage at step $t$, computed using the Generalized Advantage Estimation method with parameters $\gamma = 0.99$ and $\lambda = 0.95$. The clipping operation limits the change in $r_t(\theta)$ and stabilises the policy updates. The actor and critic networks are updated with learning rates $\alpha_\pi = 1 \times 10^{-4}$ and $\alpha_V = 5 \times 10^{-4}$ respectively. The value network parameters $\phi$ are updated by minimising the squared error loss $L^V = (V_\phi(s_t) - R)^2$.

The complete set of hyperparameters is summarised in Table~\ref{tab:ppo_params} to ensure reproducibility of the proposed approach. The training proceeds for $N_{\text{ep}} = 1500$ episodes with the batch update schedule performed every 64 episodes as described above. The overall training procedure is summarised in Algorithm~\ref{alg:ppo}. The algorithm records the best allocation vector $\mathbf{a}_{\text{best}}$ observed during training, which is used in the numerical evaluation. 

The computational complexity of the IRS allocation problem motivates the use of reinforcement learning rather than exhaustive search. For $K$ eavesdroppers and $N_{\text{irs}}$ elements, each element can be allocated to one of $K+1$ users, resulting in $(K+1)^{N_{\text{irs}}}$ possible allocations. Evaluating the secrecy capacity for each allocation requires numerical integration over the signal bandwidth, making exhaustive enumeration computationally prohibitive. As the number of eavesdroppers or IRS elements increases, the search space grows exponentially. The proposed PPO-based approach circumvents this exponential complexity by learning a policy through episodic training with $N_{\text{ep}} = 1500$ episodes, where the training complexity scales linearly as $O(N_{\text{ep}} \cdot N_{\text{irs}} \cdot (K+1))$. Once trained, the policy network generates allocations through simple forward passes with $O(1)$ complexity per inference, enabling practical reconfiguration when channel conditions change due to user mobility. The linear scaling with respect to $N_{\text{irs}}$ and $K$ makes the approach feasible even for larger systems, unlike exhaustive search, which grows exponentially.

\begin{algorithm}[t]
\label{alg:ppo}
\caption{PPO Based Secrecy Capacity Maximisation}
\begin{algorithmic}[1]
\Statex \textbf{Input.} Hyper-parameters as per Table~\ref{tab:ppo_params}. Channel gains $g^{\scriptscriptstyle \text{LOS}}_{\scriptscriptstyle k,\text{LED}}$ and $g^{\scriptscriptstyle \text{NLOS}}_{\scriptscriptstyle k,n,\text{LED}}$ for all $k$ and $n$. Fixed constants $\Lambda$, $T_s$, $K$, $N_{\text{irs}}$.
\Statex \textbf{Output.} Allocation vector $\mathbf{a}_{\text{best}}$ maximising \eqref{eq:final_cs_colluding_a}.
\State Initialise policy network $\pi_\theta$ and value network $V_\phi$.
\State Select a random initial allocation $\mathbf{a}_{\text{best}}$ and set $C_{s,\max} \leftarrow C_s^{\text{coll}}(\mathbf{a}_{\text{best}})$.
\For{episode $= 1$ to $N_{\text{ep}}$}
    \State Initialise empty trajectory buffer $\mathcal{D}$.
    \State $s_1 \leftarrow \emptyset$, $\mathbf{a} \leftarrow [\;]$.
    \For{$n = 1$ to $N_{\text{irs}}$}
        \State Sample $a_n \sim \pi_{\theta_{\text{old}}}(\cdot \mid s_n)$.
        \State Append $a_n$ to $\mathbf{a}$ and set $s_{n+1} \leftarrow (a_1,\dots,a_n)$.
    \EndFor
    \State Compute terminal reward $R = C_s^{\text{coll}}(\mathbf{a})$ using \eqref{eq:final_cs_colluding_a}.
    \State Store $(s_1, a_1, \dots, s_{N_{\text{irs}}}, a_{N_{\text{irs}}}, R)$ in $\mathcal{D}$.
    \If{$R > C_{s,\max}$}
        \State $C_{s,\max} \leftarrow R$, $\mathbf{a}_{\text{best}} \leftarrow \mathbf{a}$.
    \EndIf
    
    \If{episode $\mod N_{\text{batch}} = 0$} \Comment{Update policy after collecting batch}
        \State Compute advantage estimates $\hat{A}_t$ for all steps in $\mathcal{D}$ using GAE with $\gamma$ and $\lambda$.
        \State Update $\theta$ by maximising $L^{\text{CLIP}}(\theta)$ using samples from $\mathcal{D}$.
        \State Update $\phi$ by minimising $(V_\phi(s_t) - R)^2$.
        \State Set $\theta_{\text{old}} \leftarrow \theta$.
        \State Clear buffer $\mathcal{D} \leftarrow \emptyset$.
    \EndIf
\EndFor
\State \textbf{return} $\mathbf{a}_{\text{best}}$.
\end{algorithmic}
\end{algorithm}

\begin{table}[t]
\centering
\caption{PPO Hyper-parameters for Reproducible Training Configuration.}
\label{tab:ppo_params}
\renewcommand{\arraystretch}{1.15}
\setlength{\tabcolsep}{6pt}
\begin{tabular}{|>{\centering\arraybackslash}p{0.14\columnwidth}
                |>{\raggedright\arraybackslash}p{0.28\columnwidth}
                |>{\raggedright\arraybackslash}p{0.32\columnwidth}|}
\hline
\textbf{Symbol} & \textbf{Quantity} & \textbf{Value} \\
\hline
$N_{\text{ep}}$    & Training episodes       & 1500 \\
\hline
$N_{\text{batch}}$ & Batch size              & 64 \\
\hline
$\epsilon$         & Clipping parameter      & 0.2 \\
\hline
$\lambda$          & GAE parameter           & 0.95 \\
\hline
$\gamma$           & Discount factor         & 0.99 \\
\hline
$\alpha_{\pi}$     & Actor learning rate     & $1\times10^{-4}$ \\
\hline
$\alpha_{V}$       & Critic learning rate    & $5\times10^{-4}$ \\
\hline
$d_h$              & Hidden layer dimension  & 256 \\
\hline
\end{tabular}
\end{table}
\section{Simulation Results}

This section presents simulation results that illustrate the performance of the proposed \ac{ppo} based \ac{irs} allocation algorithm. The basic principle of the approach is to increase the received signal quality at Bob while degrading the received signal quality at the eavesdroppers by exploiting constructive and destructive interference created by the \ac{irs} induced time delays. To observe these effects, an indoor environment of dimensions $5\text{ m} \times 5\text{ m} \times 3\text{ m}$ is considered, with one \ac{led} mounted at the centre of the ceiling and an \ac{irs} attached to the wall at $x=0$. The \ac{ppo} training hyperparameters are given in Table~\ref{tab:ppo_params}. 
The main room layout and \ac{irs} configuration are summarised in 
Table~\ref{tab:sim_layout}, and the physical and system parameters are listed in 
Table~\ref{table:simulation_parameters}.

\begin{table}[t]
\centering
\caption{Room layout, user positions, and IRS configuration}
\label{tab:sim_layout}
\renewcommand{\arraystretch}{1.15}
\setlength{\tabcolsep}{6pt}
\begin{tabular}{|>{\centering\arraybackslash}p{0.16\columnwidth}
                |>{\raggedright\arraybackslash}p{0.32\columnwidth}
                |>{\raggedright\arraybackslash}p{0.33\columnwidth}|}
\hline
\textbf{Symbol} & \textbf{Quantity} & \textbf{Value} \\
\hline
-- & Room dimensions & $5 \times 5 \times 3$ m$^3$ \\
\hline
-- & LED position & $(2.5, 2.5, 3.0)$ m \\
\hline
-- & IRS position & Wall at $x=0$ \\
\hline
$N_{\text{irs}}$ & Number of IRS elements & $15 \times 15$ \\
\hline
$P_{\text{opt}}$ & Transmit optical power & $1$ W to $10$ W \\
\hline
\multicolumn{3}{|c|}{\textbf{Best case}} \\
\hline
$\mathbf{p}_{\text{B}}$ & Bob position & $(2.5, 2.5, 0.75)$ m \\
\hline
$\mathbf{p}_{\text{E}_1}$ & Eve$_1$ position & $(4.5, 4.5, 0.75)$ m \\
\hline
$\mathbf{p}_{\text{E}_2}$ & Eve$_2$ position & $(4.0, 4.0, 0.75)$ m \\
\hline
\multicolumn{3}{|c|}{\textbf{Worst case}} \\
\hline
$\mathbf{p}_{\text{B}}$ & Bob position & $(4.5, 4.5, 0.75)$ m \\
\hline
$\mathbf{p}_{\text{E}_1}$ & Eve$_1$ position & $(2.5, 2.5, 0.75)$ m \\
\hline
$\mathbf{p}_{\text{E}_2}$ & Eve$_2$ position & $(2.0, 2.0, 0.75)$ m \\
\hline
\end{tabular}
\end{table}

{\small
\begin{table}[t!]
\centering
\caption{System parameters}
\label{table:simulation_parameters}
\renewcommand{\arraystretch}{1.15}
\setlength{\tabcolsep}{6pt}
\begin{tabular}{|>{\centering\arraybackslash}p{0.16\columnwidth}
                |>{\raggedright\arraybackslash}p{0.34\columnwidth}
                |>{\raggedright\arraybackslash}p{0.33\columnwidth}|}
\hline
\textbf{Symbol} & \textbf{Quantity} & \textbf{Value} \\
\hline
$\Phi_{\scriptscriptstyle 1/2}$ & LED half-power semi-angle & $60^\circ$ \\
\hline
$R_{\scriptscriptstyle \text{pd}}$ & PD responsivity [A/W] & $0.6$ \\
\hline
$A_{\scriptscriptstyle d}$ & PD physical area [cm$^2$] & $1$ \\
\hline
$\Psi_{\scriptscriptstyle \text{FoV}}$ & PD FoV & $90^\circ$ \\
\hline
$T_s$ & Symbol period [ns] & $1$ \\
\hline
$\mathcal{N}$ & Noise power spectral density & $10^{-21}$ \\
\hline
$\rho$ & IRS reflectivity & $1$ \\
\hline
$\xi$ & Refractive index of PD lens & $1.5$ \\
\hline
$\mathsf{k}$ & Modulation scaling factor & $3.2$ \\
\hline
$\Gamma$ & Gap value [dB] & $2$ \\
\hline
$D$ & Element separation [cm] & $30$ \\
\hline
\end{tabular}
\end{table}
}

\begin{figure}[t]
\centering
\centerline{\includegraphics[width=3.5in]{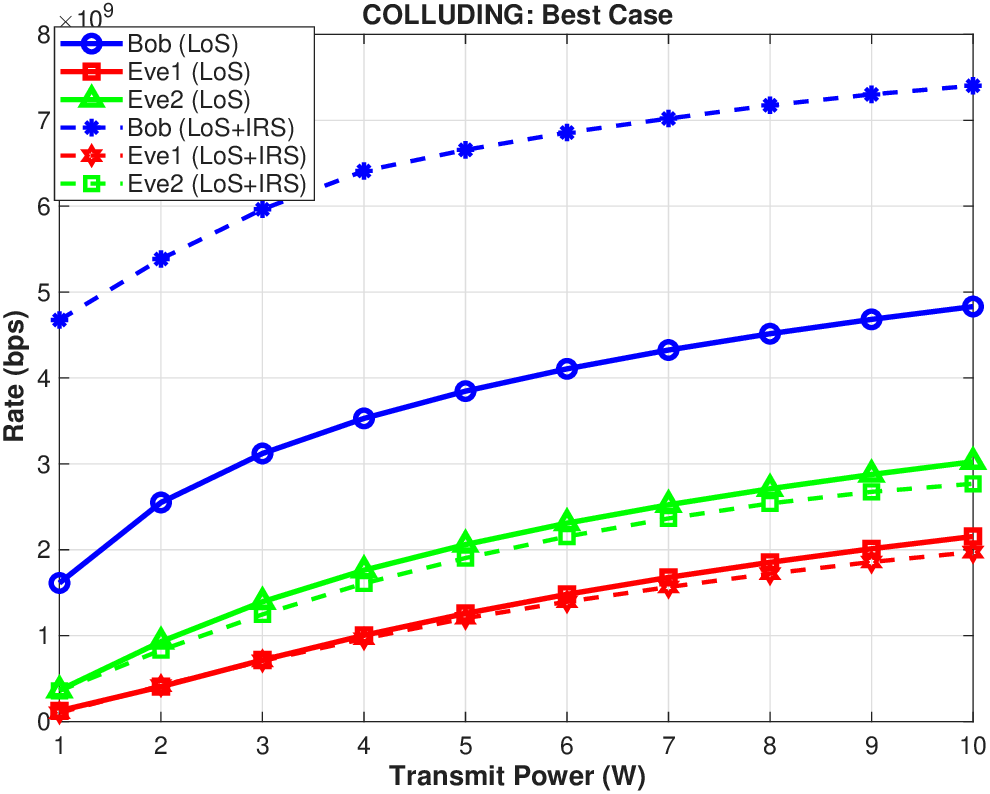}}
\caption{Achievable rates versus transmit power for the colluding scenario (best case geometry).}
\label{fig:coll_best}
\end{figure}

Fig.~\ref{fig:coll_best} shows the achievable rates of Bob, Eve$_1$, and Eve$_2$ as functions of the transmit optical power in the colluding eavesdropper scenario, for the best–case geometry where Bob is located directly beneath the LED and thus receives the strongest \ac{los} signal.
The solid curves correspond to the \ac{los}–only baseline, whereas the dashed curves represent the \ac{los}+\ac{irs} case with PPO–based allocation. It can be observed that Bob’s rate is consistently higher when the \ac{irs} is used. At $P_{\text{opt}} = 6$~W, the PPO–optimised allocation provides an increase of roughly \textbf{67\%} in Bob’s rate compared with the \ac{los}–only link. In contrast, the rates of Eve$_1$ and Eve$_2$ are slightly reduced by the optimised allocation: at $P_{\text{opt}} = 6$~W the rate of Eve$_1$ decreases by about \textbf{5.8\%}, while that of Eve$_2$ decreases by about \textbf{6.7\%} relative to their \ac{los}–only values.

\begin{figure}[!t]
\centering
\centerline{\includegraphics[width=3.5in]{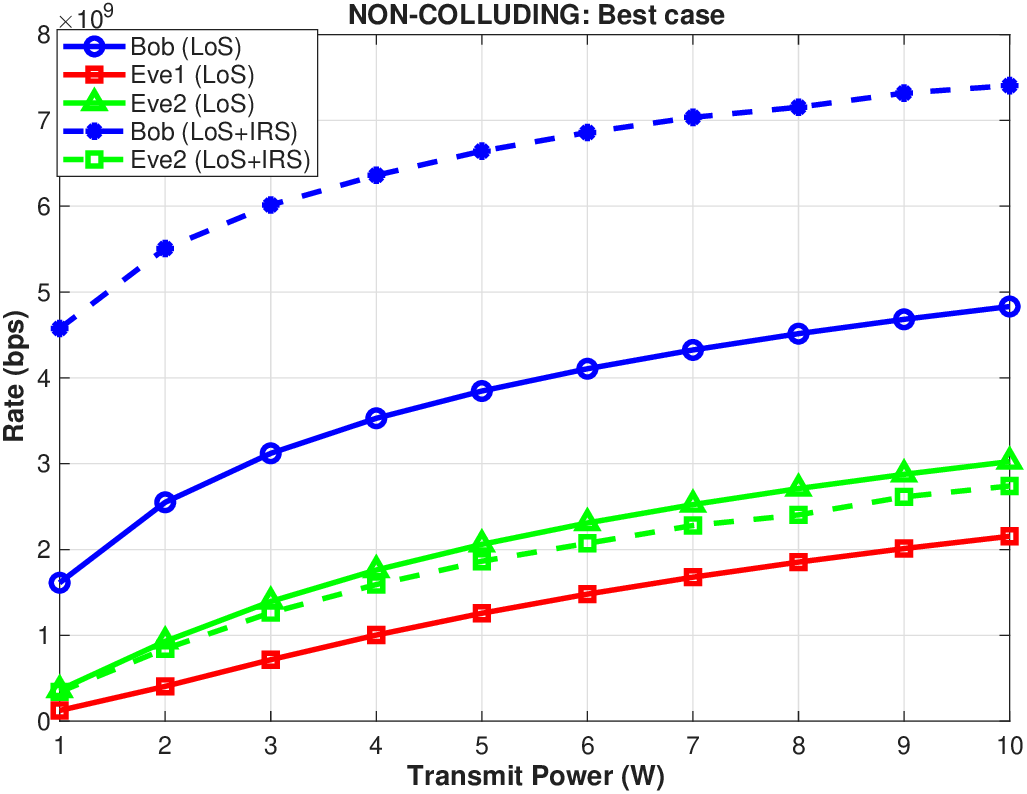}}
\caption{Achievable rates versus transmit power for the non-colluding scenario (best-case geometry).}
\label{fig:nc_best}
\end{figure}

Fig.~\ref{fig:nc_best} shows the achievable rates of Bob, Eve$_1$, and Eve$_2$ in the non-colluding eavesdropping scenario under the best-case geometry. When the \ac{irs} allocation learned by the \ac{ppo} agent is applied, Bob’s rate is significantly increased compared with the \ac{los}-only baseline; at $P_{\text{opt}} = 6$~W, the rate enhancement at Bob is approximately \textbf{67\%}. Among the eavesdroppers, Eve$_2$ is the strongest and therefore dominates the secrecy performance. With the proposed \ac{irs} allocation, Eve$_2$’s rate is reduced by about \textbf{10.3\%} at the same transmit power, while Eve$_1$’s rate remains essentially unchanged over the whole power range. In the non-colluding case the secrecy capacity is determined only by the eavesdropper with the highest rate, so the unchanged rate of Eve$_1$ does not affect the secrecy metric. This simultaneous increase in Bob’s rate and reduction in the dominant eavesdropper’s rate creates a favourable secrecy gap in the non-colluding case.

\begin{figure}[!t]
\centering
\centerline{\includegraphics[width=3.5in]{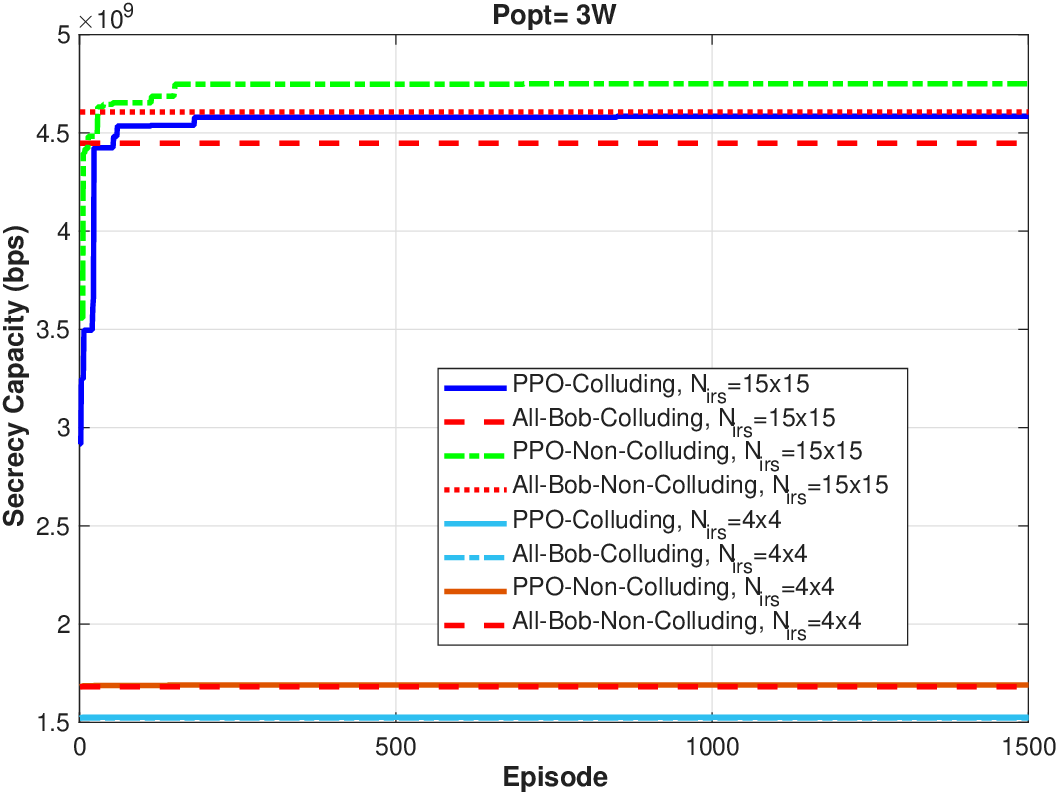}}
\caption{Secrecy capacity versus training episodes for colluding and non-colluding eavesdroppers with different IRS sizes.}
\label{fig:conv_best}
\end{figure}

\begin{figure}[!t]
\centering
\centerline{\includegraphics[width=3.5in]{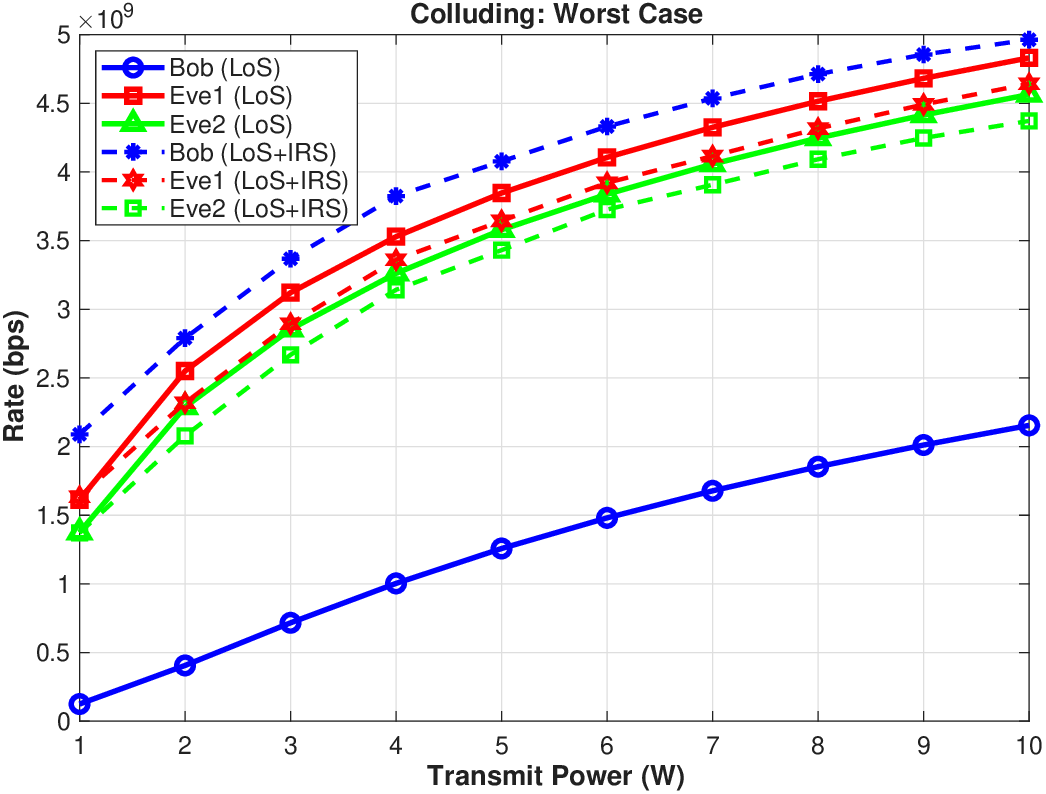}}
\caption{Secrecy capacity versus training episodes for colluding and non-colluding eavesdroppers with different IRS sizes.}
\label{fig:conv_best}
\end{figure}

Fig.~\ref{fig:conv_best} shows the secrecy capacity versus training episodes at $P_{\text{opt}} = 3$~W for colluding and non-colluding eavesdroppers with two \ac{irs} sizes, $4\times4$ and $15\times15$. The All-Bob curves, corresponding to the baseline where all IRS elements are allocated to Bob, are flat because the allocation is fixed, whereas the PPO curves increase during the first episodes and then settle to a stable value, which indicates that a good allocation policy is learned. For the $15\times15$ array, the final secrecy capacity achieved by the \ac{ppo} policy is about \textbf{3\%} higher than that of the All-Bob baseline in both scenarios. Comparing the two array sizes, the $15\times15$ surface provides higher secrecy capacity than the $4\times4$ surface, since the larger array offers more elements to shape the channel. At the same time, the $4\times4$ case converges very quickly in the first few episodes because the action space is smaller. Overall, the figure shows that the proposed \ac{ppo}-based allocation improves secrecy and converges reliably for both \ac{irs} sizes.

\begin{figure}[!t]
\centering
\centerline{\includegraphics[width=3.5in]{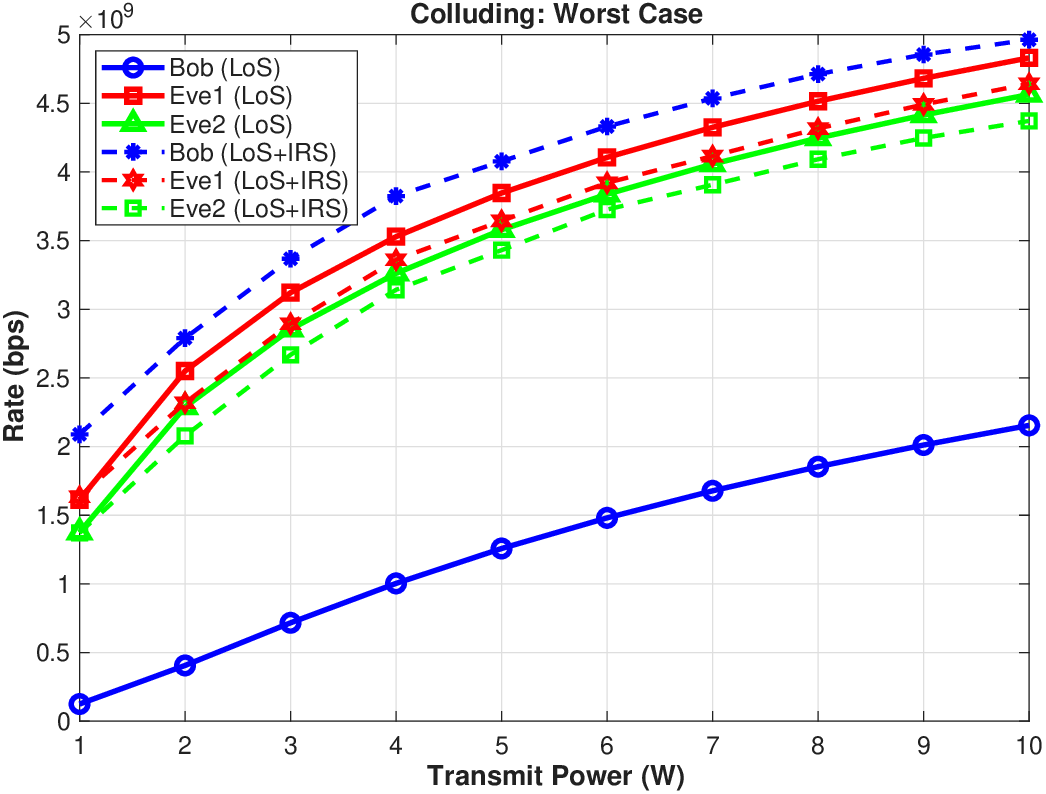}}
\caption{Achievable rates versus transmit power for the colluding worst–case geometry.}
\label{fig:coll_worst}
\end{figure}
Fig.~\ref{fig:coll_worst} shows the achievable rates of Bob, Eve$_1$, and Eve$_2$ in the colluding worst–case geometry, where Bob is located near the cell edge while both eavesdroppers are closer to the \ac{led}. In the LoS–only case, the eavesdroppers enjoy a clear advantage over Bob across the whole transmit–power range. When the \ac{irs} is optimised by the proposed \ac{ppo} algorithm, Bob’s rate is significantly boosted, with an increase of about \textbf{192.5\%} at $P_t = 6$~W compared with the LoS baseline, while Eve$_1$ and Eve$_2$ experience rate reductions of approximately \textbf{4.5\%} and \textbf{3.0\%}, respectively. These results indicate that, even under highly unfavourable geometry, the proposed \ac{irs} allocation is able to create a positive secrecy gap by strongly enhancing Bob’s link while slightly degrading the colluding eavesdroppers’ links.
\begin{figure}[!t]
\centering
\centerline{\includegraphics[width=3.5in]{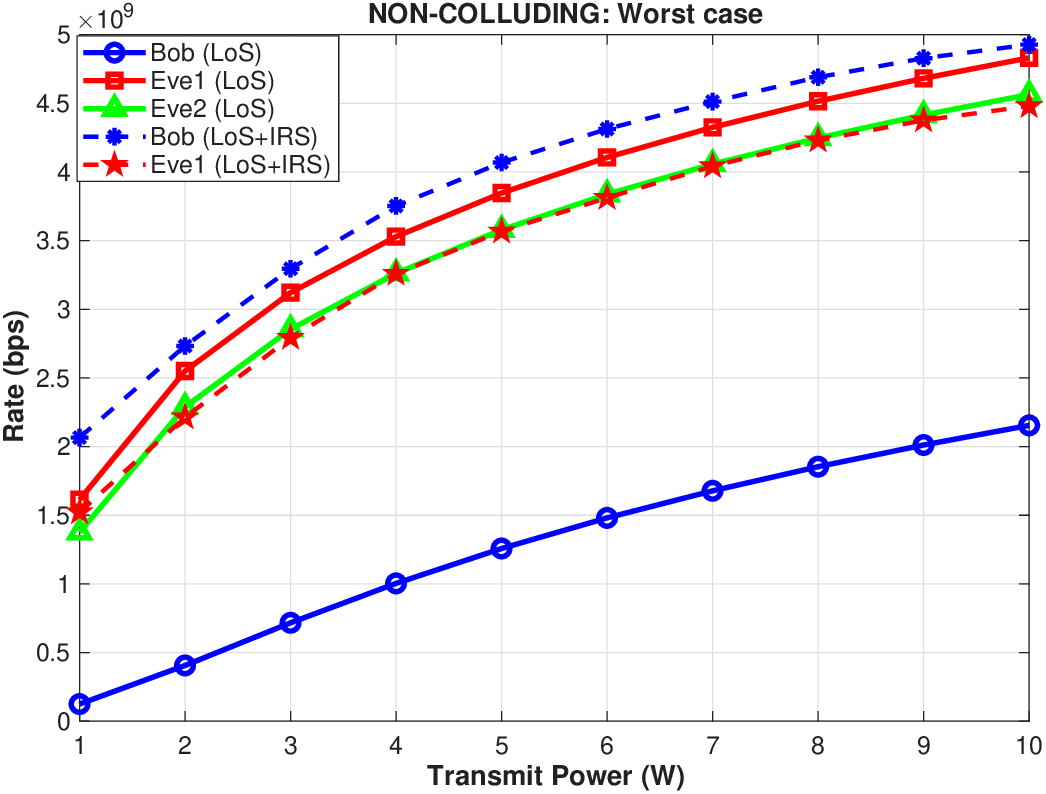}}
\caption{Achievable rates versus transmit power in the non-colluding worst-case scenario.}
\label{fig:nc_worst}
\end{figure}

Fig.~\ref{fig:nc_worst} illustrates the achievable rates for Bob and the eavesdroppers in the non-colluding worst-case scenario, where Bob is located at the cell edge while the eavesdroppers are closer to the \ac{led}. Even under this unfavourable geometry, the proposed \ac{ppo}-based \ac{irs} allocation is able to create a clear rate advantage for Bob. At a moderate transmit power of $P_t=6$~W, Bob's rate with \ac{irs} assistance is increased by approximately \textbf{191\%} compared to the \ac{los}-only case. In contrast, the rate of the dominant Eve$_1$ is reduced by about \textbf{7\%} when the \ac{irs} allocations is optimised by \ac{ppo}.

\begin{figure}[!t]
\centering
\centerline{\includegraphics[width=3.5in]{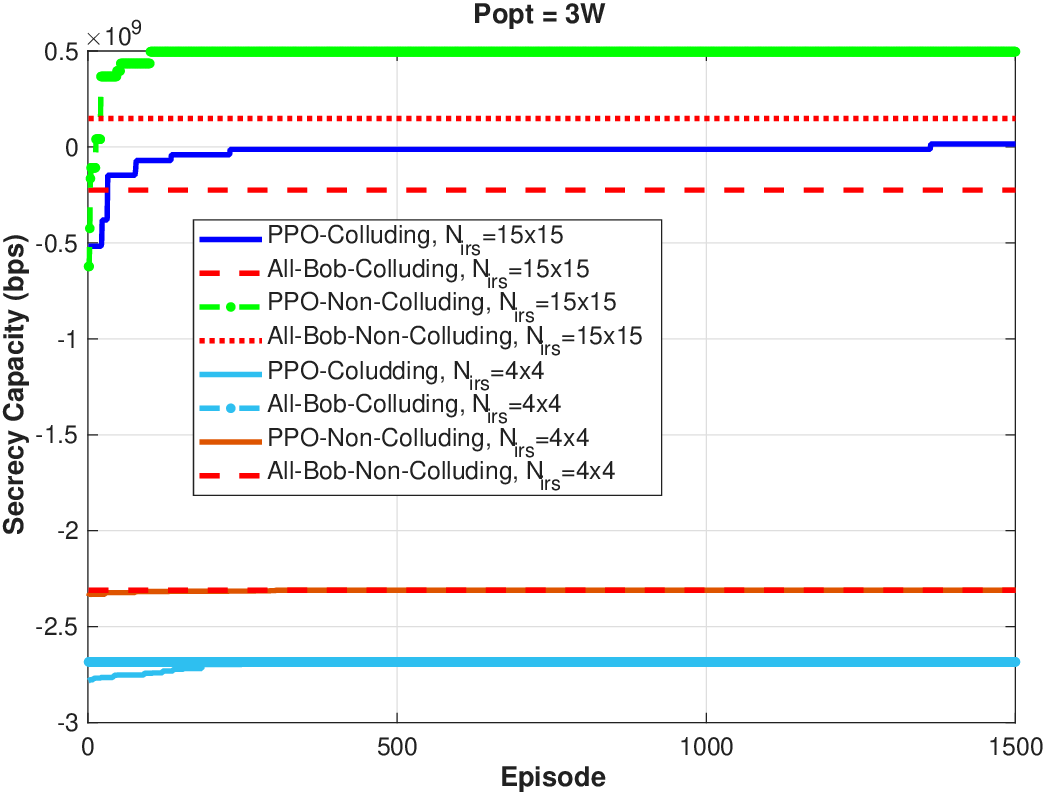}}
\caption{Secrecy capacity versus training episodes for colluding and non–colluding eavesdroppers in the worst–case geometry.}
\label{fig:conv_worst}
\end{figure}

Fig.~\ref{fig:conv_worst} shows the secrecy capacity versus training episode at $P_{\text{opt}} = 3$~W in the worst case geometry. In the colluding case, the eavesdroppers combine their received signals, so their effective \ac{snr} is very strong and the All-Bob baseline stays in the negative region throughout training. In contrast, the PPO–colluding curve starts from negative secrecy capacity but gradually increases and crosses above zero, meaning that the learned allocation is able to restore a secure link. At convergence, the secrecy capacity achieved by PPO is a little more than twice that of the All-Bob scheme, corresponding to an improvement of about \textbf{107\%}. A similar behaviour is observed in the non-colluding case. The PPO–non-colluding curve quickly rises and converges to a positive secrecy capacity that is roughly \textbf{235\%} higher than the All-Bob baseline. For the smaller $4\times4$ \ac{irs}, both cases remain in the negative region because of the limited number of elements, and the PPO-based scheme also gives a negative secrecy capacity but converges very quickly. Overall, the figure shows that the PPO agent learns an effective policy, turning a negative secrecy capacity into a positive secrecy capacity for the $15\times15$ array and consistently outperforming the All-Bob allocation.

\begin{figure}[h!]
\centering
\centerline{\includegraphics[width=3.4in]{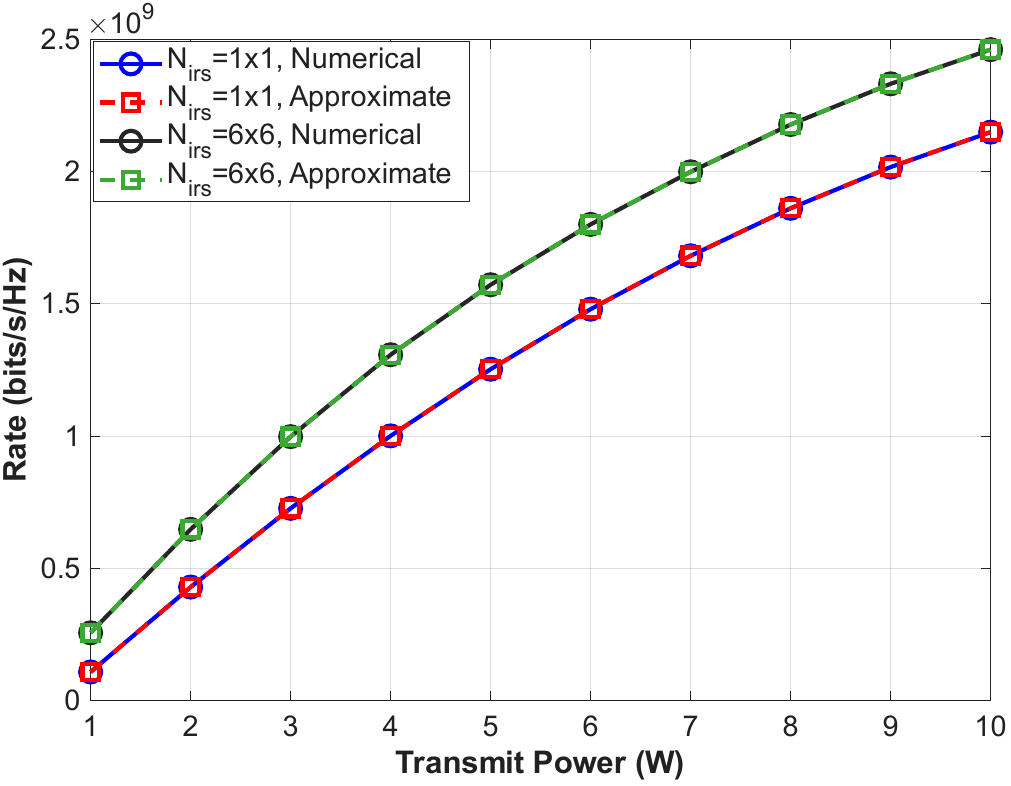}}
\caption{Validation of the analytical rate expression against exact numerical integration for different numbers of IRS elements.}
\label{Ana_Appro}
\end{figure}

Fig.~\ref{Ana_Appro} validates the closed-form approximate rate expression in \eqref{equa:approx_k} against the exact numerical integration in \eqref{rate} for two different \ac{irs} configurations. The results confirm that the analytical approximation remains highly consistent with the numerical values across the entire range of transmit power for a representative edge user. As shown in the figure, increasing the number of \ac{irs} elements results in a higher achievable rate. This close agreement between the curves demonstrates that the proposed analytical model accurately captures the channel behaviour while providing a computationally efficient alternative to numerical integration.

\section{Conclusion}
This paper proposed a novel \ac{pls} framework for indoor \ac{irs}-assisted VLC systems by exploiting the frequency-selective fading induced by reflection path delays. We formulated the secrecy capacity maximisation problem under a robust threat model involving colluding and non-colluding eavesdroppers and developed a \ac{ppo}-based \ac{rl} algorithm to solve the resulting combinatorial optimisation challenge. By learning to allocate \ac{irs} elements to engineer constructive interference at the legitimate receiver and destructive \ac{isi} at the eavesdroppers, the proposed solution significantly outperforms static baselines. Numerical results demonstrate that \ac{irs}-induced time-delay assumption effectively secures the communication link, ensuring a positive secrecy capacity even in geometrically disadvantageous scenarios where the eavesdroppers possess stronger channel conditions than the legitimate user.

\bibliographystyle{IEEEtran}
\bibliography{Refrences.bib}
\begin{IEEEbiography}[{\includegraphics[width=1in,height=1.25in,clip,keepaspectratio]{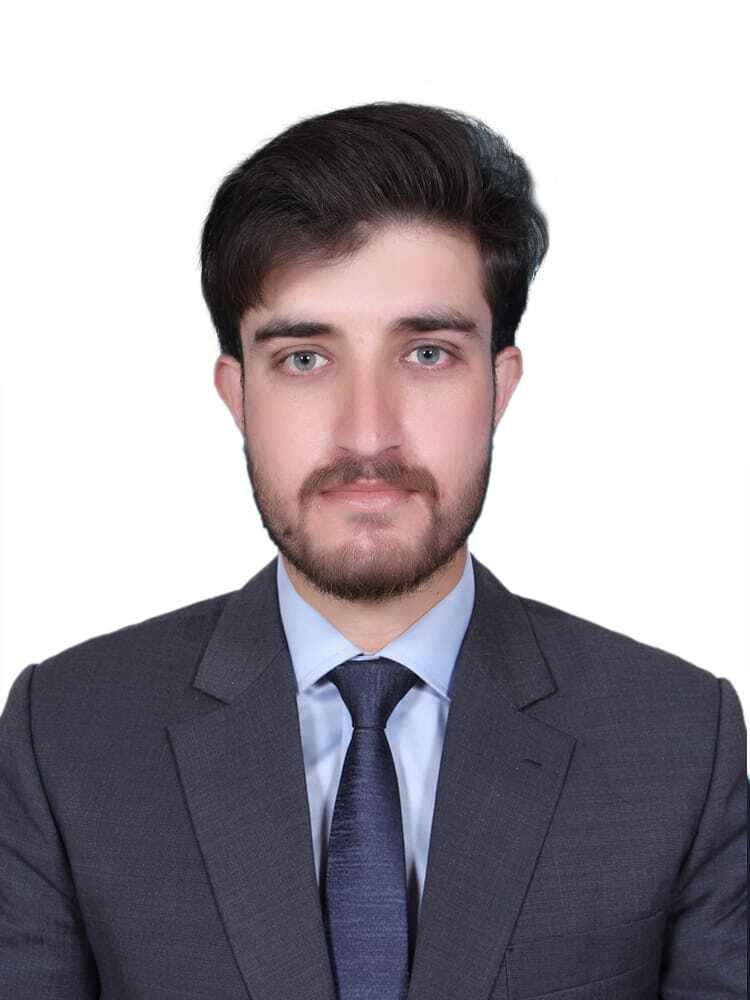}}]
{Rashid Iqbal } (Graduate Student Member, IEEE) received his B.Sc. degree in Electrical Engineering from CECOS University of IT and Emerging Sciences, Peshawar, Pakistan, in 2019. He earned his M.Sc. degree in Electrical and Electronic Engineering from the Institute of Space Technology (IST), Islamabad, Pakistan, in 2022. He is pursuing a Ph.D. in Electrical and Electronic Engineering at the University of Glasgow, UK (2023–2026). His research interests include indoor visible light communication (VLC), physical layer security (PLS), optical intelligent reflecting surfaces (OIRS), integrated sensing and communication (ISAC), channel modeling, and optimization techniques for wireless communications.
\end{IEEEbiography}

\begin{IEEEbiography}[{\includegraphics[width=1in,height=1.25in,clip,keepaspectratio]{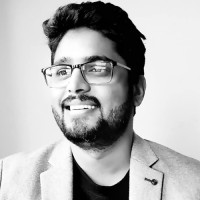}}]
{Ahmed Zoha }(Senior Member, IEEE) received the PhD degree from the esteemed 6G/5GIC Centre, University of Surrey, Guildford, U.K., in 2014. He is currently an associate professor with the James Watt School of Engineering, University of Glasgow, Glasgow, U.K. He has more than 15 years of experience. His research interests include AI, machine learning, and advanced signal processing. He was the recipient of three prestigious IEEE best paper awards. He was endorsed as an exceptional U.K. talent by the Royal Academy of Engineering and awarded to early-career world-leading innovators and scientists.
\end{IEEEbiography}

\begin{IEEEbiography}[{\includegraphics[width=1in,height=1.25in,clip,keepaspectratio]{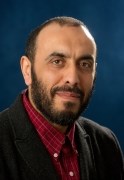}}]
{Salama Ikki } (Senior Member, IEEE) is currently a Professor and the Research Chair of wireless communications with Lakehead University, Thunder Bay, ON, Canada. He is the author of more than 100 journals and conference papers and has more than 8000 citations and an H-index of 41. His research
group has made substantial contributions to 4G and 5G wireless technologies. His group’s current research interests include massive MIMO, cell-free massive MIMO, visible light communications, and wireless sensor networks. He was a recipient of several awards for research, teaching, and services.
\end{IEEEbiography}

\begin{IEEEbiography}[{\includegraphics[width=1in,height=1.25in,clip,keepaspectratio]{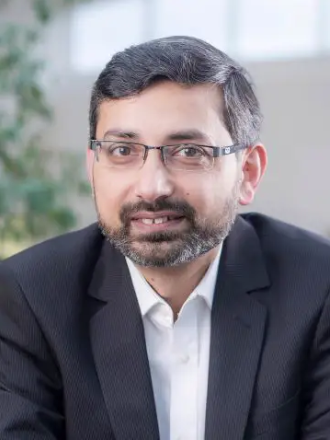}}]
{MOHAMMAD Ali IMRAN } (Fellow, IEEE) is a Professor of Communication Systems, Dean of Transnational Engineering Education, Dean of Graduate Studies for College of Science and Engineering and head of the Communications, Sensing and Imaging hub at the University of Glasgow,
UK. He also holds affiliate professor positions with the University of Oklahoma, USA, and the 5G/6G Innovation Centre of the University of Surrey, UK. He has a grant portfolio of \pounds 20M+, published 600 journal and conference papers, and edited/published 13 books. He is a Fellow of IEEE, the Royal Society of Arts, the Royal Society of Edinburgh, the European Alliance of Innovation, and the Institution of Engineering and Technology.
\end{IEEEbiography}

\begin{IEEEbiography}[{\includegraphics[width=1in,height=1.25in,clip,keepaspectratio]{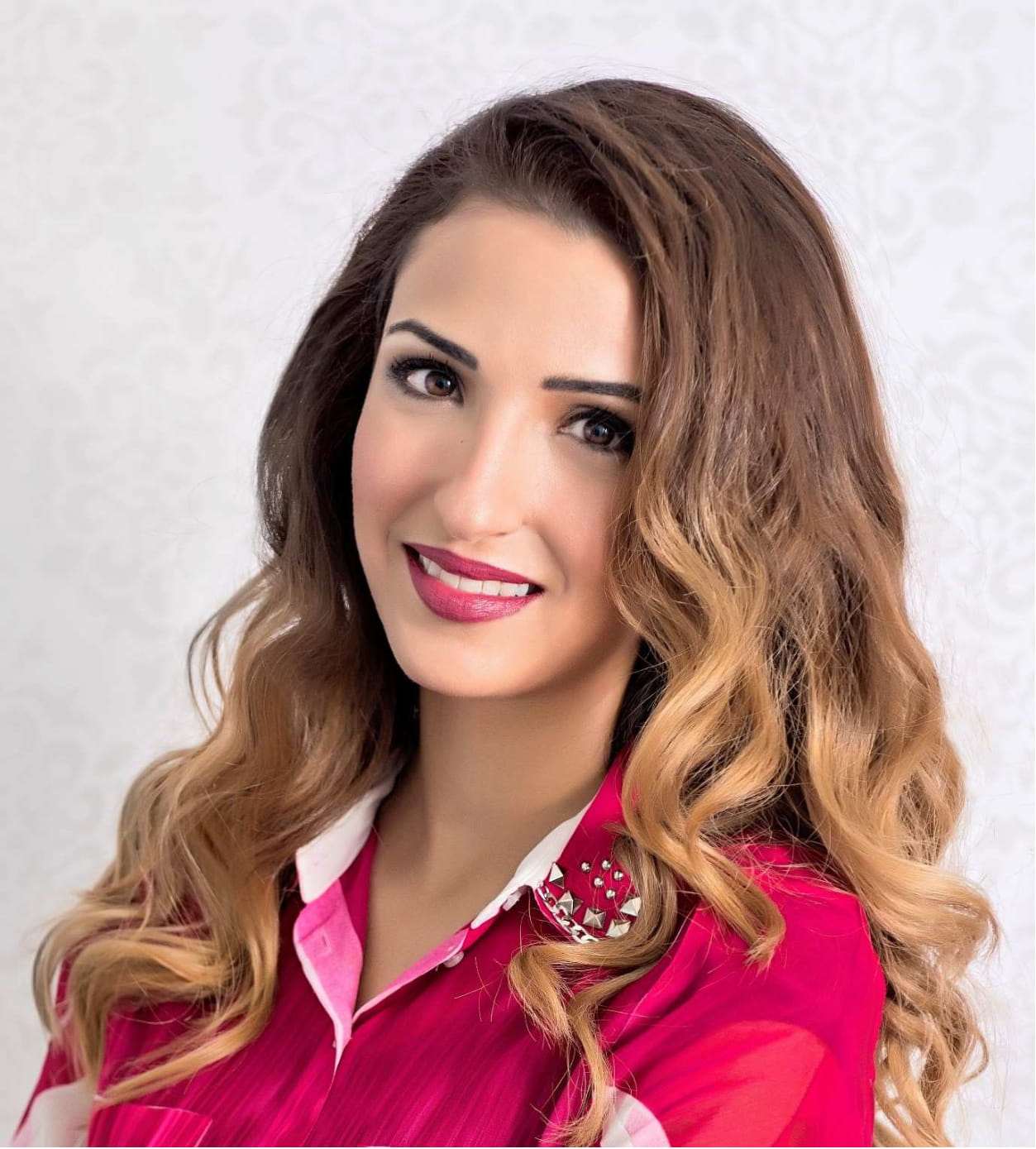}}]
{Dr Hanaa Abumarshoud }  (Senior Member, IEEE) is a Lecturer (Assistant Professor) in the James Watt School of Engineering at the University of Glasgow, UK. Prior to this, she was a Postdoctoral Research Associate at the LiFi Research and Development Centre at the University of Strathclyde (2020–2022) and the University of Edinburgh (2017–2020). She received her MSc and PhD in Electrical and Computer Engineering from Khalifa University, UAE, in 2013 and 2017, respectively. Dr Abumarshoud was endorsed as a “Global Talent” by the Royal Academy of Engineering in the UK in 2022 and was named among the “100 Brilliant and Inspiring Women in 6G” by the Women in 6G community in both 2024 and 2025 for her contributions to the field. She is a Senior Member of IEEE and served as an Associate Editor for IEEE Communications Letters (2022-2024). She regularly contributes as an Organising Committee Member for flagship conferences, organises and chairs technical workshops, and serves as a reviewer for leading international journals and conferences within IEEE, the Royal Society, and the Optical Society. She received the IEEE TAOS Best Paper Award in 2022.
\end{IEEEbiography}

\end{document}